\documentclass[superscriptaddress,prd,twocolumn]{revtex4}
\usepackage{latexsym}
\usepackage{amsthm}
\usepackage{amsmath}
\usepackage{graphicx}
\usepackage{amssymb}
\usepackage{bbm}

\newcommand{\ind}{\hspace*{\parindent}}

\def\OO{{\cal O}}

\def\epm{e^+e^-}

\newcommand{\GeV}{\,\mathrm{GeV}}
\newcommand{\MeV}{\,\mathrm{MeV}}

\newcommand{\be}{\begin{equation}}
\newcommand{\ee}{\end{equation}}
\newcommand{\bea}{\begin{eqnarray}}
\newcommand{\eea}{\end{eqnarray}}
\newcommand{\f}[2]{\ensuremath{\frac{#1}{#2}}}
\newcommand{\bef}{\begin{figure}[htbp]\begin{center}}
\newcommand{\eef}{\end{center}\end{figure}}

\newcommand{\gsim}{\lower.7ex\hbox{$\;\stackrel{\textstyle>}{\sim}\;$}}
\newcommand{\lsim}{\lower.7ex\hbox{$\;\stackrel{\textstyle<}{\sim}\;$}}

\newcommand{\de}{\ensuremath $^{\circ}$}
\newcommand{\muA}{\ensuremath $\mu A$}
\newcommand{\sq}{\ensuremath $\qquad\!\!\!$}

\begin{document}

\pagestyle{plain}

\title{
\begin{flushright}
\mbox{\normalsize SLAC-PUB-13882}\\
\mbox{\normalsize SU-ITP-10/01}
\end{flushright}
\Large
An Electron Fixed Target Experiment to Search \\ 
for a New Vector Boson $A'$ Decaying to $e^+e^-$}
\author{Rouven Essig}
\affiliation{Theory Group, SLAC National Accelerator Laboratory, Menlo Park, CA 94025}
\author{Philip Schuster}
\affiliation{Theory Group, SLAC National Accelerator Laboratory, Menlo Park, CA 94025}
\author{Natalia Toro}
\affiliation{Theory Group, Stanford University, Stanford, CA 94305}
\author{Bogdan Wojtsekhowski}
\affiliation{Thomas Jefferson National Accelerator Facility, Newport News, VA 23606}
\date{\today}
\begin{abstract}

We describe an experiment to search for a new vector boson $A'$
with weak coupling $\alpha' \gtrsim 6\times 10^{-8} \alpha$ to electrons
($\alpha=e^2/4\pi$) in the mass range 65 MeV $<m_{A'} < $ 550 MeV.
New vector bosons with such small couplings arise naturally from a
small kinetic mixing of the ``dark photon'' $A'$ with the photon ---
one of the very few ways in which new forces can couple to the
Standard Model --- and have received considerable attention as
an explanation of various dark matter related anomalies. $A'$ bosons are
produced by radiation off an electron beam, and could appear as narrow
resonances with small production cross-section in the trident $e^+e^-$
spectrum.  We summarize the experimental approach described in a proposal 
submitted to Jefferson Laboratory's PAC35, PR-10-009 \cite{proposal}.  
This experiment, the \emph{$A'$ Experiment} (APEX), uses the electron beam of the 
Continuous Electron Beam Accelerator Facility at Jefferson Laboratory (CEBAF) at energies of 
$\approx$ 1--4 GeV incident on $0.5-10\%$ radiation length 
Tungsten wire mesh targets, and measures the resulting
$e^+e^-$ pairs to search for the $A'$ using the High Resolution 
Spectrometer and the septum magnet in Hall A.  With a $\sim 1$ month run,
APEX will achieve very good sensitivity because
the statistics of $e^+e^-$ pairs will be $\sim 10,000$ times larger
in the explored mass range than any previous search for the $A'$ boson.  
These statistics and the
excellent mass resolution of the spectrometers allow sensitivity to
$\alpha'/\alpha$ one to three orders of magnitude below current
limits, in a region of parameter space of great theoretical and
phenomenological interest. 
Similar experiments could also be performed at 
other facilities, such as the Mainz Microtron.  
\end{abstract}
\maketitle

\section{Introduction}\label{sec:introduction}
The development of the Standard Model of particle interactions is the culmination of a century of searches and 
analyses with fixed-target and colliding beam experiments. Interactions with new forces beyond the Standard Model 
are currently limited by well-tested gauge symmetries to a handful of possibilities. One of the few remaining ways 
for interactions with new sub-GeV vector-like forces to arise is for charged particles to 
acquire millicharges, $\epsilon e$, under these forces. This occurs through a simple and generic mechanism 
proposed by Holdom \cite{Holdom:1985ag}, in which a new vector particle $A_\mu^\prime$ mixes via 
quantum loops with the Standard Model photon. MeV--GeV masses for the $A'$ 
gauge boson are particularly well-motivated in this context. Such sub-GeV forces are a common feature of 
extensions of the Standard Model, but existing constraints are surprisingly weak, with limits at 
$\epsilon e\lesssim (0.3-1) \times 10^{-2}e$.

Fixed-target experiments with high-intensity electron beams and existing precision spectrometers are 
ideally suited to explore sub-GeV forces by probing reactions in which a new $A'$ vector 
particle is produced by radiation off an electron beam \cite{Bjorken:2009mm,Reece:2009un}.  The $A'$ 
can decay to an electron and positron pair and appears as a narrow resonance of small magnitude 
in the invariant mass spectrum.  The production rate of $A's$, the luminosity, and the 
mass resolution attainable at, for example, Jefferson 
Laboratory (JLab), the Mainz Microtron, and the SLAC National Accelerator Laboratory 
vastly exceeds what is currently available using colliding electron beam facilities. 
In \cite{Bjorken:2009mm}, several fixed-target experimental strategies were outlined to search 
for new sub-GeV vector interactions. 
In this paper, we summarize a concrete $A'$ search using Jefferson Laboratory's Continuous Electron 
Beam Accelerator Facility (CEBAF) and the High Resolution Spectrometers (HRS) in Hall A \cite{proposal}, 
highlighting the features that are applicable to similar experimental facilities.
This experiment, the \emph{$A'$ Experiment} (APEX), can probe charged particle 
couplings with new forces as small as 
$2\times 10^{-4} e$ and masses between $65~\MeV$ and $550~\MeV$ --- an improvement 
by more than two orders of magnitude in cross section sensitivity over all previous experiments.

Fixed-target experiments of this form are particularly timely in light of a series of recent anomalies from terrestrial, balloon-borne, and satellite experiments that suggest that
dark matter interacts with Standard Model particles. Much of this data sharply hints that dark matter is directly charged under a new force mediated by an $A'$ and not described by the Standard Model. Theoretical as well as phenomenological expectations suggest an $A'$ mass $m_{A'}\lesssim 1 \GeV$ and $\epsilon e \lesssim 10^{-2} e$.

In this paper, we shall focus on a search for new vector bosons.  However, it should be emphasized that 
this experiment will provide a powerful probe for \emph{any} new particle 
--- vector, pseudo-vector, scalar, or pseudo-scalar --- 
that has sub-GeV mass and couples to electrons 
(for other collider, accelerator, and direct and indirect astrophysical probes see \cite{Pospelov:2008jd,Batell:2009yf,Essig:2009nc,Batell:2009di,Bossi:2009uw,Yin:2009mc,Essig:2010a,Freytsis:2009bh,Baumgart:2009tn,Cheung:2009su,Abazov:2009hn,Batell:2009jf,Batell:2009zp,Schuster:2009au,Schuster:2009fc,Meade:2009mu,Yin:2009yt,Essig:2009jx,Galli:2009zc,Slatyer:2009yq,Essig:2010}; a proposal for an electron beam 
incident on a diffuse Hydrogen gas target using the Jefferson Laboratory's Free Electron Laser has been discussed 
in \cite{Freytsis:2009bh}).
 
\subsection{Brief overview of the experimental strategy}

The goal of the experiment is to measure the invariant mass spectrum
of electron-positron pairs produced by electron scattering on a
high-$Z$ target, and search for a narrow peak with width corresponding
to the instrumental resolution. The electron and positron are detected
in magnetic spectrometers with acceptance over a small range of
particle momentum and angle, such that each experimental setting is
sensitive to a mass window $\sim \pm 30\%$ about a central mass
value.  Using four beam energies from 1--4 GeV, APEX will scan the
$e^+e^-$ spectrum in the mass range $65~\MeV$ to $550~\MeV$.  

Optimal sensitivity for these masses is achieved by studying symmetric
$e^+e^-$ kinematics, where each particle carries approximately half
the beam energy and has an opening angle $\approx 5\deg$ relative to
the beam.  Such small effective angles for the spectrometer can be
achieved using a septum magnet~\cite{PREX,proposal}.  Without a septum
magnet, lower beam energies and correspondingly wider angles could be
used to probe the same mass range.  The impact of the geometry on the
physics reach will be reviewed in \S \ref{sec:reaction} and was
discussed in detail in \cite{Bjorken:2009mm}.

The experimental sensitivity is determined by statistics and mass resolution.  Given the precision of spectrometers used, the latter is limited by multiple
scattering in the target material.  In APEX, a long, tilted wire mesh target is used to obtain excellent relative mass resolution of 0.5\%.  In addition, different segments of the target will enter the spectrometers for different central angles, increasing the size of mass window probed by each setting.

With a beam of 80~$\mu$A on 0.5\%--10\% radiation-length targets at
various beam energies, we expect to collect true coincidence $e^+e^-$
events with a rate in the range 100--500 Hz (the expected background
and accidental coincidence rates within a 2~ns timing window are about
an order of magnitude lower).  The total $e^+ e^-$ sample size will
exceed $10^8$ pairs in a 6-day period for each setting, or a 12-day
period for the 4 GeV setting.

While this paper reflects an experimental setup optimized for the
equipment in Hall A at JLab, many of the experimental considerations
are also applicable for equipment available at the Mainz Microtron,
JLab Hall B, and other experimental facilities.

\subsection{Expected reach and impact}
APEX will be sensitive to new gauge bosons with couplings as small as 
$\alpha'/\alpha \sim (6-8)\times 10^{-8} $ for masses in the range $65-300 \MeV$, 
and couplings as small as $\alpha'/\alpha\sim 2 \times 10^{-7}$ for larger $m_{A'} \lesssim 525$ MeV.  
This is about a factor of 
$\sim 3-35$ times lower in $\epsilon$ than existing constraints (which assume that the 
$A'$ couples also to muons), and corresponds to 
$\sim 10-1000$ times smaller cross-sections. 

The precise mass range probed by this type of experiment can be varied by changing 
the spectrometer angular settings and/or the beam energies.  
Thus, other experimental facilities may be able to perform experiments 
similar to APEX, but targeting complementary regions of parameter space. 

The parameter range probed by APEX is interesting for several reasons.  
This region of mass and coupling is compatible
with $A'$'s explaining the annual modulation signal seen by the dark matter direct detection 
experiment DAMA/LIBRA, and also with dark matter annihilating into $A'$'s, which explains 
a myriad of recent cosmic-ray and other astrophysical anomalies (see \S
\ref{subsec:motivation}).  In addition, and independently of any connection to dark matter, 
the proposed experiment would be the
first to probe $A'$s of mass $\gtrsim 50 \MeV$ with gauge kinetic mixing below 
$\epsilon \sim 10^{-3}$, the range most compatible 
if the Standard Model hypercharge gauge force is part of a Grand Unified Theory.  

The importance for fundamental physics of discovering new forces near the GeV scale 
cannot be overstated.  

\subsection{The organization of this paper}

The paper is organized as follows. 
In \S \ref{sec:physics}, we present the physics of hypothetical
$A'$ particles, motivation for their existence, current limits, and estimated sensitivity for \emph{potential} future analyses of existing data.  
In \S \ref{sec:reaction}, we describe $A'$ production in fixed-target experiments. 
In \S \ref{sec:exp_setup}, we describe the experimental setup.  
In \S\ref{sec:signalTrident}, we present the parametrics and the Monte Carlo (MC) 
simulations of the QED $e^+e^-$ pair production rate and 
the $A'$ signal rate in the proposed setup.  We also describe how we made the sensitivity plots.  
Other background rates, such as $\pi^+$ or $e^+$ singles and accidental $e^+e^-$ pairs, 
are discussed in \S \ref{sec:backgrounds}.
The expected sensitivity is discussed in \S \ref{sec:proposed}.
The paper is summarized in \S \ref{sec:conclusion}.
Three appendices discuss the form factors used to calculate the signal and background rates 
(\S\ref{sec:chi}), the mass resolution (\S\ref{sec:resolution}), and the validation of the rates we obtain 
with the various MC simulations (\S \ref{sec:validation}).  

\section{Physics}\label{sec:physics}
\begin{figure}[tbh!]
\begin{center}
\includegraphics[width=0.47\textwidth]{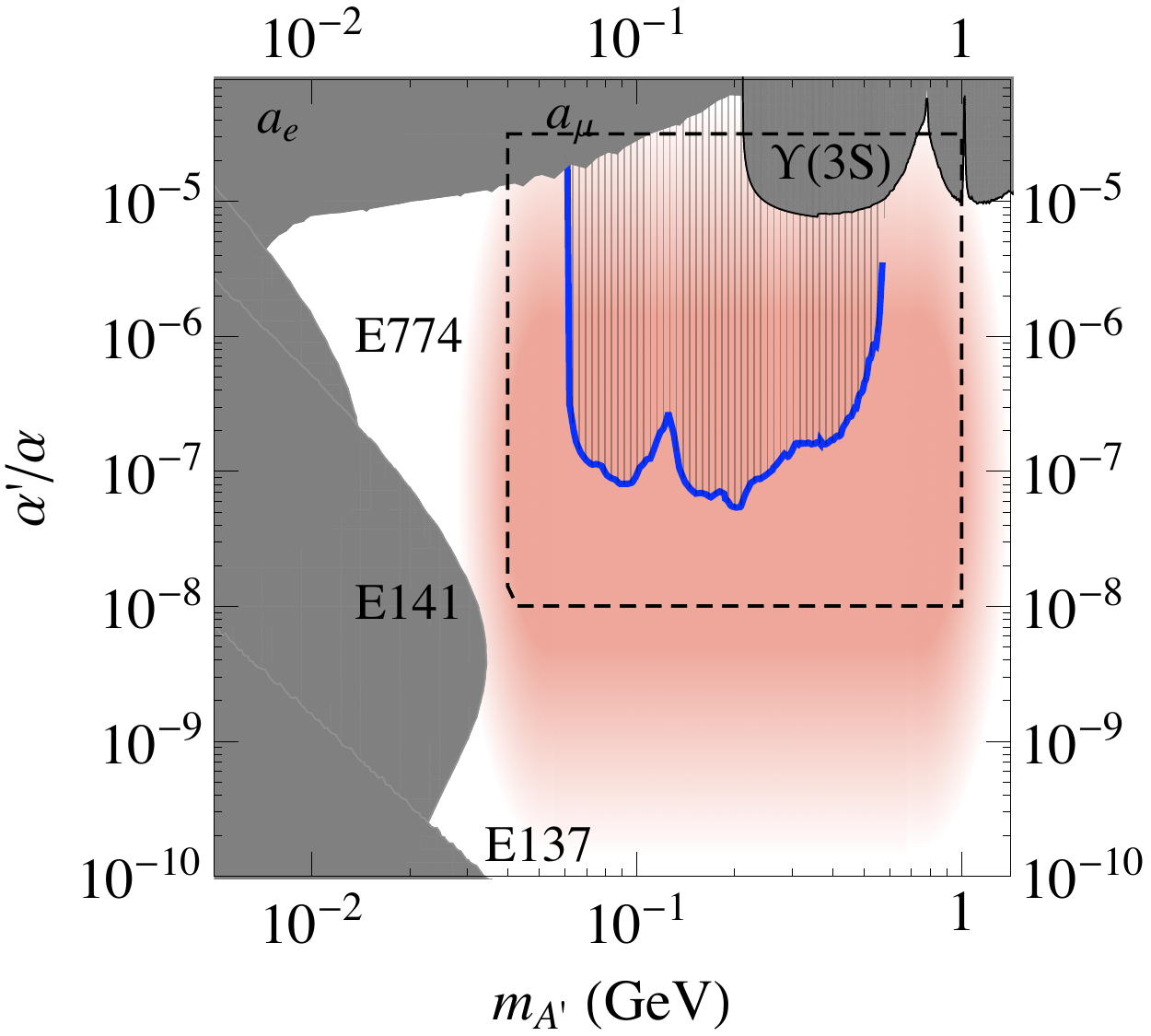}
\end{center}
\caption{Anticipated 2$\sigma$ sensitivity in $\alpha'/\alpha = \epsilon^2$ 
   for the $A'$ experiment (APEX) at Hall A in JLab (thick blue line), with existing constraints
  on an $A'$ from electron and muon anomalous magnetic moment
  measurements, $a_e$ and $a_{\mu}$ (see \cite{Pospelov:2008zw}), the
  BaBar search for $\Upsilon(3S)\to \gamma\mu^+\mu^-$ \cite{:2009cp},
  and three beam dump experiments, E137, E141, and E774
  \cite{Bjorken:1988as,Riordan:1987aw,Bross:1989mp} (see
  \cite{Bjorken:2009mm}).  The $a_{\mu}$ and $\Upsilon(3S)$ limits
  assume equal-strength couplings to electrons and muons.  The red
  region indicates the region of greatest theoretical interest, as
  described in the text.  The gray dashed line indicates the scale
  used for other plots in this paper. 
  The irregularity of the reach is an artifact of combining several different 
  run settings (see Table \ref{tab:bigBGtable}).  
  The precise mass range probed by this type of experiment can be 
  varied by changing the spectrometer angular settings and/or the beam energies.  
  We stress this point as other experimental facilities may be able to 
  perform experiments similar to APEX, but targeting complementary 
  regions of parameter space. 
\label{fig:bigSummary}}
\end{figure}

We consider new sub-GeV mass vector bosons --- `dark photons' 
$A'$ --- that couple very weakly to electrons (as mentioned previously, similar considerations apply to pseudo-vectors, scalars, and pseudo-scalars 
with sub-GeV mass that couple to electrons).  It is useful to
parameterize the coupling $g'$ of the $A'$ to electrons by a
dimensionless $\epsilon \equiv g'/e$, where $e$ is the electron
charge.  Cross-sections for $A'$ production then scale as
$\alpha'/\alpha = \epsilon^2$, where $\alpha'= g'^2/(4\pi)$ and
$\alpha=e^2/(4\pi)$ are the fine-structure constants for the dark photon and ordinary
electromagnetic interactions, respectively.  This experiment will search for 
$A'$ bosons with mass  $m_{A'} \sim 65$ MeV -- 550 MeV and $\alpha'/\alpha \gtrsim 6\times 10^{-8}$, which can be produced by a reaction
analogous to photon bremsstrahlung (see \S
\ref{sec:reaction}) and decays promptly to $e^+e^-$ or other charged
particle pairs. We refer the reader to Figure \ref{fig:bigSummary} for a summary of the reach of this 
experiment.

\subsection{Motivation for New Physics Near the GeV Scale}
New light vector particles, matter states, and their associated interactions are ubiquitous in
extensions of the Standard Model
\cite{Holdom:1985ag,Strassler:2006qa,Strassler:2006im,Strassler:2008bv,Dienes:1996zr,Abel:2003ue,Abel:2008ai,Ringwald:2008cu,Nomura:2008ru,Mardon:2009gw}.
However, the symmetries of the Standard Model restrict the interaction
of ordinary matter with such new states.  Indeed, most interactions
consistent with Standard Model gauge symmetries and Lorentz
invariance have couplings suppressed by a high mass scale.  One of the
few unsuppressed interactions is the coupling of charged Standard Model particles $\psi$
\be 
\delta {\cal L}  = g' A'_\mu \bar \psi\gamma^\mu \psi 
\ee 
to a new gauge boson $A'$, which
is quite poorly constrained for small $g'$ (see Figure
\ref{fig:bigSummary})\cite{Bjorken:2009mm}.  Similar couplings between
the $A'$ and other Standard Model fermions are also allowed, with
relations between their couplings (anomaly cancellation) required for
the $A'$ gauge symmetry to be quantum-mechanically consistent.  For
example, the $A'$ can couple only to electrons and muons, with opposite
charges $g'_e = - g'_\mu$ ( a $U(1)_{e-\mu}$  boson), or can have
couplings proportional to the electromagnetic charges $q_i$ of each fermion,
$g_i = \epsilon e q_i$.

$A'$ couplings to Standard Model matter with the latter structure can
be \emph{induced} by ordinary electromagnetic interactions through the
kinetic mixing interaction proposed by Holdom~\cite{Holdom:1985ag},
\be\label{kmix} \delta {\cal L} = \frac{\epsilon_Y}{2} F'_{\mu\nu}
F^{\mu\nu}_Y, \ee where
$F'_{\mu\nu}=\partial_{\mu}A'_{\nu}-\partial_{\nu}A'_{\mu}$ is the
field strength of the $A'$ gauge boson, and similarly $F^{\mu\nu}_Y$
is the hypercharge field strength.  This effect is generic, ensures
that the $A'$ interactions respect parity, and (as we discuss below)
naturally produces small $g'$ and $A'$ masses near the GeV scale.
This mixing is equivalent in low-energy interactions to assigning a
charge $\epsilon e q_i$ to Standard Model particles of electromagnetic
charge $q_i$, where $\epsilon = \epsilon_Y/(\cos\theta_W)$ and
$\theta_W$ is the Weinberg mixing angle.  The $A'$ couplings to
neutrinos and parity-violating couplings are negligible compared to
$Z$-mediated effects (see e.g.~\cite{Baumgart:2009tn}).

As noted in \cite{Holdom:1985ag}, a new gauge boson $A'$ that does not
couple to Standard Model matter at a classical level can still couple
through quantum-mechanical corrections.  For example, loops of any 
particle $X$ that couples to both the $A'$ and Standard
Model hypercharge generates mixing of the form \eqref{kmix}, with
\be
\epsilon \sim 10^{-3}-10^{-2} \qquad (\alpha'/\alpha \sim 10^{-6}-10^{-4}).
\ee
These quantum effects are
significant regardless of the mass $m_X$ of the particle in
question, which could be well above the TeV scale (or even at the
Planck scale) and thus evade detection. 

Smaller $\epsilon$ are expected if
nature has enhanced symmetry at high energies.  For example, it has
been conjectured that the strong and electroweak
gauge groups of the Standard Model are embedded in a grand unified
theory (GUT) with gauge group $SU(5)$ or larger that is broken
spontaneously at a high scale $M_G \approx 10^{16}$ GeV.  In this case
the mixing \eqref{kmix} is suppressed,
\be
\epsilon_{GUT} \sim \frac{\alpha_i^2}{16\pi^2}\ln{(M_{G}/M_{X})} \sim 10^{-5}-10^{-3},
\ee
where $\alpha_i$ are gauge couplings.  $\epsilon$ of this size leads
to effective couplings
\be
\alpha'/\alpha \sim 10^{-8}-10^{-6}.
\ee
As shown in Figure \ref{fig:bigSummary}, \emph{no experiment to date
  has probed the range of $\epsilon$ expected in grand unified
  theories for $m_{A'} \gtrsim 50$ MeV}.  
(From string theory, the possible range of $\epsilon$ is much larger, 
$\sim 10^{-23}-10^{-2}$
\cite{Dienes:1996zr,Abel:2003ue,Abel:2008ai,Ringwald:2008cu}.)

An $A'$ mass near but beneath the weak scale is particularly
well-motivated, as $U(1)'$ symmetry-breaking and the resulting $A'$
mass may be determined by the same physics that generates the $W$ and
$Z$ masses~\cite{ArkaniHamed:2008qp}.  The best candidate for the
origin of the weak scale is low-energy supersymmetry.  In this case,
the $A'$ can naturally acquire mass suppressed by a loop factor or
by $\sqrt{\epsilon}$ compared to the weak scale, leading to MeV to
GeV-scale $A'$ masses
\cite{Dienes:1996zr,ArkaniHamed:2008qp,Cheung:2009qd,
  Morrissey:2009ur,Katz:2009qq,Baumgart:2009tn}.  In supersymmetric
models, the gauge kinetic mixing \eqref{kmix} is accompanied by
quartic interactions \be \delta {\cal L} \sim \frac{\epsilon_Y}{4} g_Y
g_D |\phi_D|^2 |h|^2,\label{quartic} \ee between the Standard Model
Higgs doublet $h$ and any scalar $\phi_D$ charged under $U(1)'$, where
$g_Y$ and $g_D$ are the gauge couplings of Standard Model hypercharge
and the $A'$ coupling to $\phi_D$, respectively.  Electroweak symmetry
breaking gives $h$ a weak-scale vacuum expectation value, so that
\eqref{quartic} generates a mass term for $\phi_D$.  For positively
charged $\phi_D$, and sufficiently small bare mass, this mass term is negative and triggers
$U(1)'$ breaking by the Higgs mechanism. The resulting induced mass for the $A'$ is \be
\label{eq:DtermLine} 
m_{A'} \sim \sqrt{\epsilon} \sqrt\frac{g_D g_Y}{g_2^2}\,m_W \sim
\mbox{MeV--GeV}, 
\ee 
where $g_2$ is Standard Model $SU(2)_L$ gauge coupling and $m_W$ is
the W-boson mass.  The resulting mass is precisely in the $50-1000$
MeV range targeted by this experiment.  Given our $\epsilon$
sensitivity, we expect to probe the portion of this parameter space
with small $g_D$.  For example, for $g_D\sim 0.04$ and $\epsilon\sim
5\times 10^{-4}$ ($\alpha'/\alpha \sim 2.5 \times 10^{-7}$), we have
$m_{A'}\sim 400$ MeV, which can definitively be probed by the proposed
experiment. Note that the mechanism of $U(1)'$ breaking above does not
rely on supersymmetry, as any quartic interaction of the form (\ref{quartic}), with arbitrary coupling, 
can transmit electro-weak masses to the $A'$. Thus, the mass relation (\ref{eq:DtermLine})
should not be interpreted too literally. 

We stress that the mass of the $A'$ breaks any apparent symmetry
between it and the photon: though Standard Model particles have
induced $\epsilon$-suppressed charges under the $A'$, any new matter
charged under the $A'$ would \emph{not} have any effective coupling to
the photon, and would have gone undetected. 

An electron beam scattering on a high-$Z$ target such as
Tungsten will produce $A'$'s through bremsstrahlung reactions with a
cross-section
\bea
\sigma_{A'} &\sim& 100 \ \mbox{pb} \left (\frac{\epsilon}{10^{-4}} 
\right )^2 \left ( \frac{100 \MeV}{m_{A'}} \right )^2,
\eea
several orders of magnitude larger than in colliding electron and
hadron beams \cite{Essig:2009nc}.  The $A'$ can decay to electrons,
and is therefore visible as a narrow resonance in the trident $e^+e^-$
mass spectrum.  

Such a new gauge boson would constitute the first discovery of a new
gauge force since the observation of $Z$-mediated neutral currents.
Besides the obvious physical interest of a fifth force, the $A'$ like
the $Z$ could open up a new ``sector'' of light, weakly coupled
particles whose spectrum and properties could be measured in
fixed-target experiments and flavor factories.  The $A'$ sector would
provide a new laboratory for many physical questions, and would be
revealing precisely because its interactions with Standard Model
particles are so weak.  In particular, if nature is approximately
supersymmetric near the TeV scale, the mass scale of supersymmetry
breaking for the $A'$ sector is naturally suppressed by $\epsilon$
times gauge couplings.  In this case, supersymmetry could be studied
easily in the $A'$ sector, and possibly even discovered there by
relatively low-energy experiments before Standard Model superpartners
are seen at colliders.

\subsection{Motivation for an $A'$ from Dark Matter}\label{subsec:motivation}
Dark matter interpretations of recent astrophysical and terrestrial
anomalies provide an urgent impetus to search for $A'$'s in the mass range 
50 MeV -- 1 GeV, with a coupling $\epsilon \sim 10^{-4} - 10^{-2}$.  

The concordance model of big bang cosmology --- the ``Lambda Cold Dark Matter'' ($\Lambda$CDM) model --- 
explains all observations of the cosmic microwave background, large scale structure formation, and supernovae, 
see e.g.~\cite{Komatsu:2008hk,Eisenstein:2005su,Perlmutter:1998np,Riess:1998cb,Kowalski:2008ez}.   
This model suggests that Standard Model particles make up only 
about 4\% of the energy density in the Universe, while ``dark energy'' and ``dark matter'' make up 74\% and 22\%, respectively, 
of the Universe's energy density.  
The concordance model does not require dark matter to have any new interactions beyond gravity 
with Standard Model particles.  
However, an intriguing theoretical observation, dubbed the \emph{WIMP miracle}, 
suggests that dark matter \emph{does} have new interactions.  
In particular, if dark matter consists of $\sim$100 GeV to 10 TeV particles interacting via the electroweak force (``weakly interacting 
massive particles'' or ``WIMPs''), they would automatically have the right relic abundance observed today.  

In addition to the WIMP miracle, evidence from cosmic-ray data and the terrestrial direct dark matter detection 
experiment DAMA/LIBRA strongly suggest that dark matter interacts with ordinary matter \emph{not} just gravitationally.  
While the WIMP miracle suggests that dark matter is charged under the Standard Model electroweak force, 
we will see that \emph{these observations provide impressive evidence for dark matter interacting with ordinary matter through 
a new force, mediated by a new 50 MeV -- 1 GeV mass gauge boson}.  
In addition to explaining any or all of these observations, dark matter charged under this new force automatically has the correct thermal 
relic abundance observed today by virtue of its interactions via the new force carrier, reproducing the success of the WIMP dark matter hypothesis.  

The satellites PAMELA \cite{Adriani:2008zr} and Fermi \cite{Abdo:2009zk}, the balloon-borne 
detector ATIC \cite{2008zzr}, the ground-based telescope HESS \cite{Collaboration:2008aaa,Aharonian:2009ah}, 
as well as other experiments, observe an excess in the cosmic-ray 
flux of electrons and/or positrons above backgrounds expected from normal astrophysical processes.    
If their source is dark matter annihilation or decay, synchrotron radiation from these electrons and positrons
could also explain the ``WMAP haze'' near the Galactic center \cite{Finkbeiner:2003im}, which consists of 
an excess seen in the WMAP Cosmic Microwave Background data.  In addition, starlight near the 
Galactic center would inverse Compton scatter off the high energy electrons and positrons and produce an 
excess in gamma-rays.  A detection of a gamma-ray excess towards the Galactic center region in the gamma-ray 
data obtained with the Fermi satellite was recently reported in \cite{Dobler:2009xz}, and has been 
dubbed the ``Fermi haze''.

\begin{figure*}
\begin{center}
\includegraphics[width=0.35\textwidth]{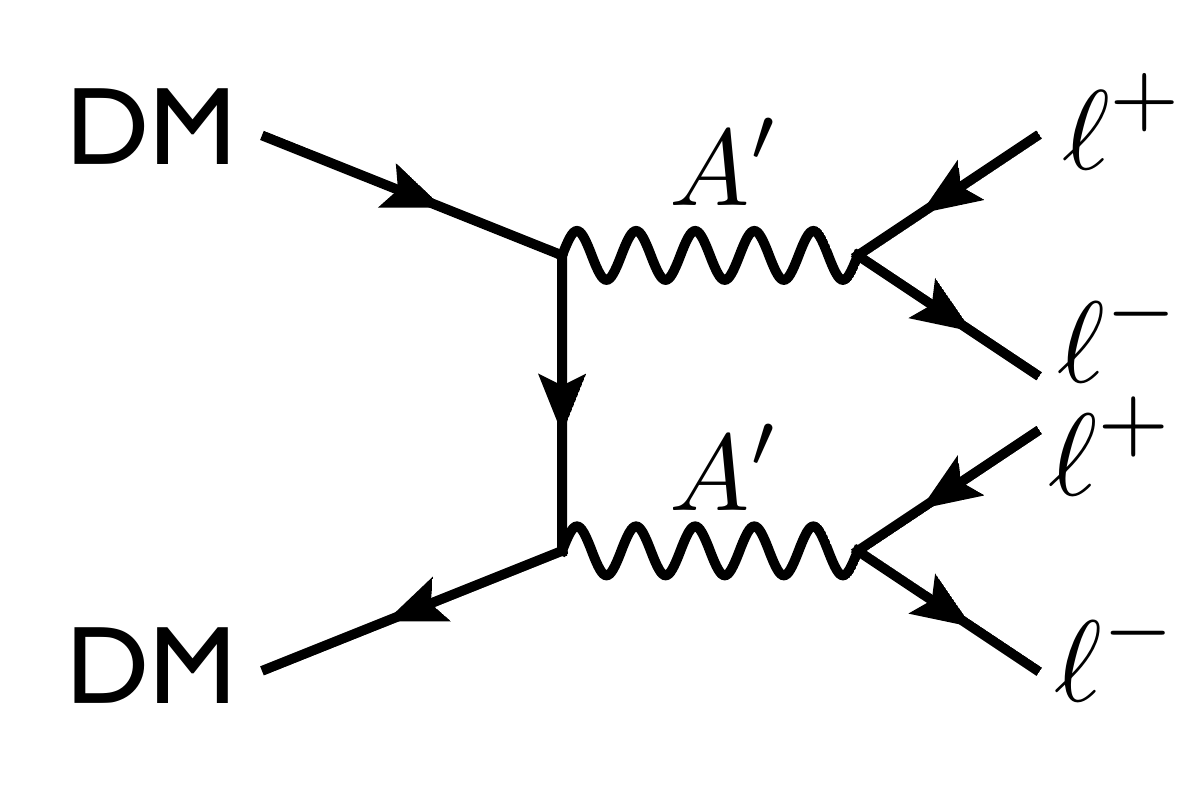}
\hspace{1.5cm} 
\includegraphics[width=0.35\textwidth]{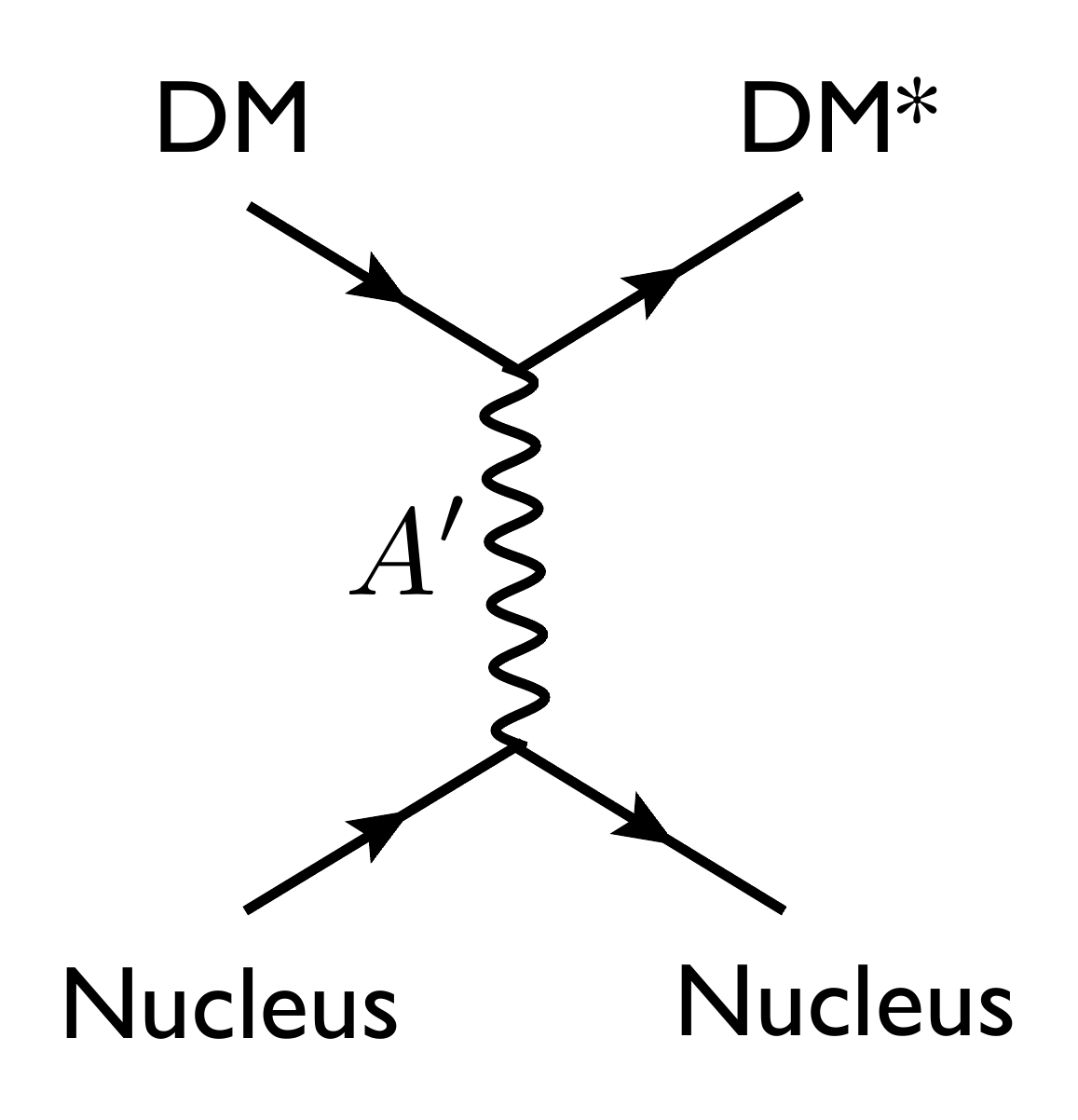}
\end{center}
\caption{{\bf Left:} Dark matter annihilation into the dark photon $A'$, which decays into charged leptons 
such as electrons and/or muons, can explain the cosmic-ray electron and/or positron excesses seen by 
PAMELA, Fermi, ATIC, HESS, and other experiments.  
{\bf Right:} Dark matter scattering into an excited state off nuclei through $A'$ exchange 
in direct dark matter detection experiments can explain the annual modulation signal observed 
by DAMA/LIBRA, and the null results of other direct detection experiments.  
\label{fig:darkmatter}}
\end{figure*}

Taken together, these observations by several experimental collaborations provide compelling evidence that there is 
an unexplained excess in cosmic-ray electrons and positrons in our Galaxy.  
Given the firm evidence for a 22\% dark matter content of the Universe, a very natural source of these excesses is dark matter annihilation.  
However, two features of these observations are incompatible with annihilation of ordinary 
thermal WIMP dark matter.   
They instead provide impressive evidence that dark matter is charged under a new $U(1)'$ 
and annihilating into the $A'$, which decays directly into electrons and positrons, or into muons that decay 
into electrons and positrons, see Figure \ref{fig:darkmatter} (left)
(see e.g.~\cite{ArkaniHamed:2008qn,Pospelov:2008jd,Hisano:2003ec,MarchRussell:2008yu,
Cirelli:2008pk,Cholis:2008wq,Cholis:2008qq,Cui:2009xq}).  
These two features are:
\begin{itemize}
\item The annihilation cross-section required to explain the signal is
  50-1000 times larger than the thermal freeze-out cross-section for an ordinary WIMP that is needed to 
  reproduce the observed dark matter relic density.  
  This can be explained if dark matter interacts with a new long range force mediated by an $\OO$(GeV) 
  mass gauge boson, which allows the dark matter annihilation cross-section ($\langle \sigma v \rangle$) 
  to be enhanced at low dark matter velocities, i.e.~$\langle \sigma v \rangle \propto 1/v$.  
  In this case, in the early Universe when the dark matter velocity was high ($\sim 0.3 c$), the annihilation cross-section 
  that determines the relic abundance can naturally be the same as that of an ordinary WIMP and reproduce
  the WIMP miracle.  However, in the Milky Way halo now, the dark matter has a much lower velocity ($v\sim 10^{-3}c$),  
  leading to a large increase in the annihilation cross-section that is required to explain the cosmic-ray data.
  The enhancement at low velocities through a new long-range force is very well known and called the Sommerfeld 
  effect \cite{Sommerfeld:1931}.  
\item
  The PAMELA satellite did \emph{not} see an anti-proton excess \cite{Adriani:2008zq}, which 
  strongly suggests that dark matter annihilation is dominantly producing leptons, and not baryons. 
  If dark matter is interacting via a $\OO(GeV)$ mass force particle in order to have a large annihilation
  rate via the Sommerfeld mechanism, then annihilations into the force carrier automatically fail to produce any baryons. 
  Kinematically, the force carriers cannot decay into baryons, and are instead forced to decay into the lighter charged leptons.
  Thus, annihilation products of dark matter are leptonic in this case.   
\end{itemize} 

To explain the additional sources of evidence for a new $\GeV$ scale force, we briefly summarize the 
consequence for dark matter mass spectra that follow from dark matter carrying a charge
under a new force. 
If dark matter is charged under a non-Abelian force that acquires mass, then radiative effects can split all components of the dark matter with size,
$\delta \sim \alpha_D \Delta m_{W_D}$, where $\alpha_D$ is the non-Abelian fine structure constant and $\Delta m_{W_D}$ is the splitting 
of gauge boson masses \cite{ArkaniHamed:2008qn}. Typically, these splittings are
$\Delta m_{W_D} \sim \alpha_D m_{W_D} \sim 1-10 \MeV$ for $m_{W_D}\sim 1 \GeV$ \cite{ArkaniHamed:2008qn}.
Thus,  $\delta\sim 100$ keV for $\alpha_D\sim 10^{-2}$. These splittings are completely analogous to the splittings 
that arise between the $\pi^{\pm}$ and $\pi^0$ from Standard Model SU(2) breaking. 
If instead a non-Abelian force confines at a scale
$\Lambda_D\sim \GeV$, then a heavy-flavor meson can be cosmologically long-lived and
thus a dark matter candidate \cite{Schuster:2009}. Hyperfine interactions
can naturally induce $\sim 100$ keV splittings of the dark matter particles in this case.   
We emphasize that the $\GeV$ scale force carrier particles mediate quantum corrections that generate 
the 100 keV and 1-10 MeV splittings of dark matter states \cite{ArkaniHamed:2008qn,TuckerSmith:2001hy,Chang:2008gd,Schuster:2009}.

When mass splittings arise, $A'$ mediated interactions of dark matter with ordinary matter as well as dark matter self-interactions are dominated by inelastic collisions \cite{ArkaniHamed:2008qn}. The direct dark matter detection experiment DAMA/LIBRA as well as the INTEGRAL telescope provide intriguing evidence for such interactions. The DAMA/NaI \cite{Bernabei:2005hj} and DAMA/LIBRA \cite{Bernabei:2008yi} experiments 
have reported an annual modulation signal over nearly eleven years of operation with more than $8\sigma$ significance.  
Modulation is expected because the Earth's velocity with respect to the dark matter halo varies as 
the Earth moves around the sun, and the phase of the observed modulation is consistent with this origin.  
A simple hypothesis that explains the spectrum and magnitude of the signal, and reconciles it with the null results of other experiments, 
is that dark matter-nucleus scattering is dominated by an inelastic process, \be \chi\;N \rightarrow \chi^* \;N, \ee in 
which the dark matter $\chi$ scatters off a nucleus $N$ into an excited state $\chi^*$ with 
mass splitting $\delta \approx 100$ keV \cite{TuckerSmith:2001hy}. The kinematics of these reactions is also remarkably consistent with all the distinctive properties of the nuclear recoil spectrum reported by DAMA/LIBRA.  In addition, the INTEGRAL telescope \cite{Strong:2005zx} has reported a 511keV photon signal near the galactic center,
 indicating a new source of $\sim$ 1-10 MeV electrons and positrons.  This excess could be explained by collisions  of  $\OO$(100 GeV-1 TeV) mass dark matter into $\OO$(MeV) excited states in the galaxy \cite{Finkbeiner:2007kk} --- dark matter excited by scattering decays back to the ground state by emitting a soft $e^+e^-$ pair. The 511keV excess then arises from the subsequent annihilation of the produced positrons. 

The existence of an $A'$ may also help explain various other particle physics 
anomalies \cite{Pospelov:2008zw} such as the anomalous magnetic moment of 
the muon ($(g-2)_\mu$) \cite{Bennett:2006fi} and the HyperCP anomaly \cite{Park:2005eka}.  

While these experimental hints provide an urgent motivation to look
for an $A'$, it is important to emphasize the value of these searches
in general.  There has never been a systematic search for new
GeV-scale force carriers that are weakly coupled to Standard Model
particles.  Nothing forbids their existence, and their discovery would
have profound implications for our understanding of nature.  A
relatively simple experiment using the facilities available at, for example, 
Jefferson Laboratory and Mainz will probe a large and interesting range of 
$A'$ masses and couplings.

\subsection{Current Limits on Light $U(1)$ Gauge Bosons}

Constraints on new $A'$'s that decay to $e^+e^-$ and the search reach of an experiment using 
the spectrometers of Hall A at Jefferson Laboratory are summarized in 
Figure \ref{fig:bigSummary}.
Shown are constraints from electron and muon anomalous magnetic moment measurements, 
$a_e$ and $a_{\mu}$ \cite{Pospelov:2008zw}, the BaBar search for 
$\Upsilon(3S)\to \gamma A' \to \gamma\mu^+\mu^-$, and 
three beam dump experiments, E137, E141, and E774 \cite{Bjorken:2009mm}.   
The constraints from $a_\mu$ and the BaBar search assume that the $A'$
couples to muons --- this is the case, for example, if it mixes with the photon.  
If it only couples to electrons, then the constraints on $\alpha'/\alpha$ and $m_{A'}$ 
in the region to which the proposed experiment is sensitive are weaker than $\alpha'/\alpha \lesssim 10^{-4}$. 
 
We refer the reader to \cite{Bjorken:2009mm,Pospelov:2008zw} for
details on existing constraints. Here, we briefly review the
constraint on $e^+e^- \to \gamma A' \to \gamma\mu^+\mu^-$ derived from
the BaBar search \cite{Aubert:2009cp}.  If the $A'$ couples to both
electrons and muons, this is the most relevant constraint in the
region probed by the proposed experiment.  The analysis of
\cite{Aubert:2009cp} was in fact a search for $\Upsilon(3S)$ decays
into a pseudoscalar $a$, $\Upsilon(3S)\to \gamma a \to
\gamma\mu^+\mu^-$, but can be interpreted as a limit on $A'$
production because the final states are identical.  Using
$\mathcal{L}_{\rm int} \sim 30$ fb$^{-1}$ of data containing $\sim
122\times 10^6$ $\Upsilon(3S)$ events, a 90\% C.L.~upper limit of
roughly $(1-4)\times 10^{-6}$ on the $\gamma \mu^+\mu^-$ branching
fraction was found for $m_{A'} \sim 2 m_\mu - 1$ GeV.  This search
would thus be sensitive to about $\sim 100-500$ events with $e^+e^-
\to \gamma A' \to \gamma\mu^+\mu^-$.  Requiring that $\sigma(e^+e^-\to
\gamma A') \times BR(A'\to \mu^+\mu^-) \times \mathcal{L}_{\rm int}
\lesssim 500$, where $BR(A'\to \mu^+\mu^-) = 1/(2+R(m_{A'}))$ for
$m_{A'} > 2m_\mu$ with $R=\f{\sigma(e^+e^- \rightarrow\,{\rm
    hadrons};\, E=m_{A'})} {\sigma(e^+e^- \rightarrow \mu^+\mu^-;\,
  E=m_{A'})}$, and rescaling the resulting constraint to represent a
95\% C.L.~upper bound, we find the constraint depicted in Figure
\ref{fig:bigSummary}.  For $m_{A'} \gtrsim 2 m_\mu$, this requires
$\alpha'/\alpha\gtrsim 10^{-5}$, while the constraint weakens at
higher masses, especially near the $\rho$-resonance.  See 
\footnote{Note that 
our estimate of the constraint here disagrees with those previously published 
in \cite{Essig:2009nc,Bjorken:2009mm} and \cite{Reece:2009un}.  
Compared to the published versions of \cite{Essig:2009nc,Bjorken:2009mm},
we have here included $R$ and also corrected an error which made the old  
estimates in \cite{Essig:2009nc,Bjorken:2009mm} too optimistic.  
The estimate of \cite{Reece:2009un} is also too optimistic, since it did not 
include the signal efficiency from using one-sigma (mass resolution) 
bin-widths, and it also used an overly optimistic mass resolution for BaBar.}
for a comparison of our sensitivity estimate to those previously published.

We caution that systematic uncertainties in the $A'$ limit beyond
those quoted in \cite{Aubert:2009cp} may slightly weaken the resulting
limit, which should therefore be taken as a rough approximation unless
further analysis is done.  First, $A'$ production in B-factories is
more forward-peaked than the $\Upsilon(3S)$ decay mode considered in
\cite{Aubert:2009cp}, so that the signal acceptance is more uncertain.
In addition, background distributions in \cite{Aubert:2009cp} are
derived from smooth polynomial fits to data collected on the
$\Upsilon(4S)$ resonance, which is assumed to contain no signal. This
assumption is not correct for $A'$ production, though the resulting
systematic effects are expected to be small.

\subsection{Sensitivity of Potential Searches using Existing Data}
\label{ssec:limits} 
Several past and current experiments have data that could be used to
significantly improve current limits on $\alpha'/\alpha$, as discussed in
\cite{Pospelov:2008zw,Reece:2009un,Batell:2009di}. 
Here, we estimate the potential
sensitivity of searches in three channels ($\pi^0 \rightarrow \gamma
A' \rightarrow \gamma e^+e^-$, $\phi \rightarrow \eta A' \rightarrow
\eta e^+e^-$, and $e^+e^- \rightarrow \gamma A' \rightarrow \gamma
\mu^+ \mu^-$), considering only the statistical uncertainties and
irreducible backgrounds.  These are likely overestimates, as we are
unable to include either systematic uncertainties or significant
instrumental backgrounds such as photon conversion in the detector
volume.

\begin{figure}
\begin{center}
\includegraphics[width=0.45\textwidth]{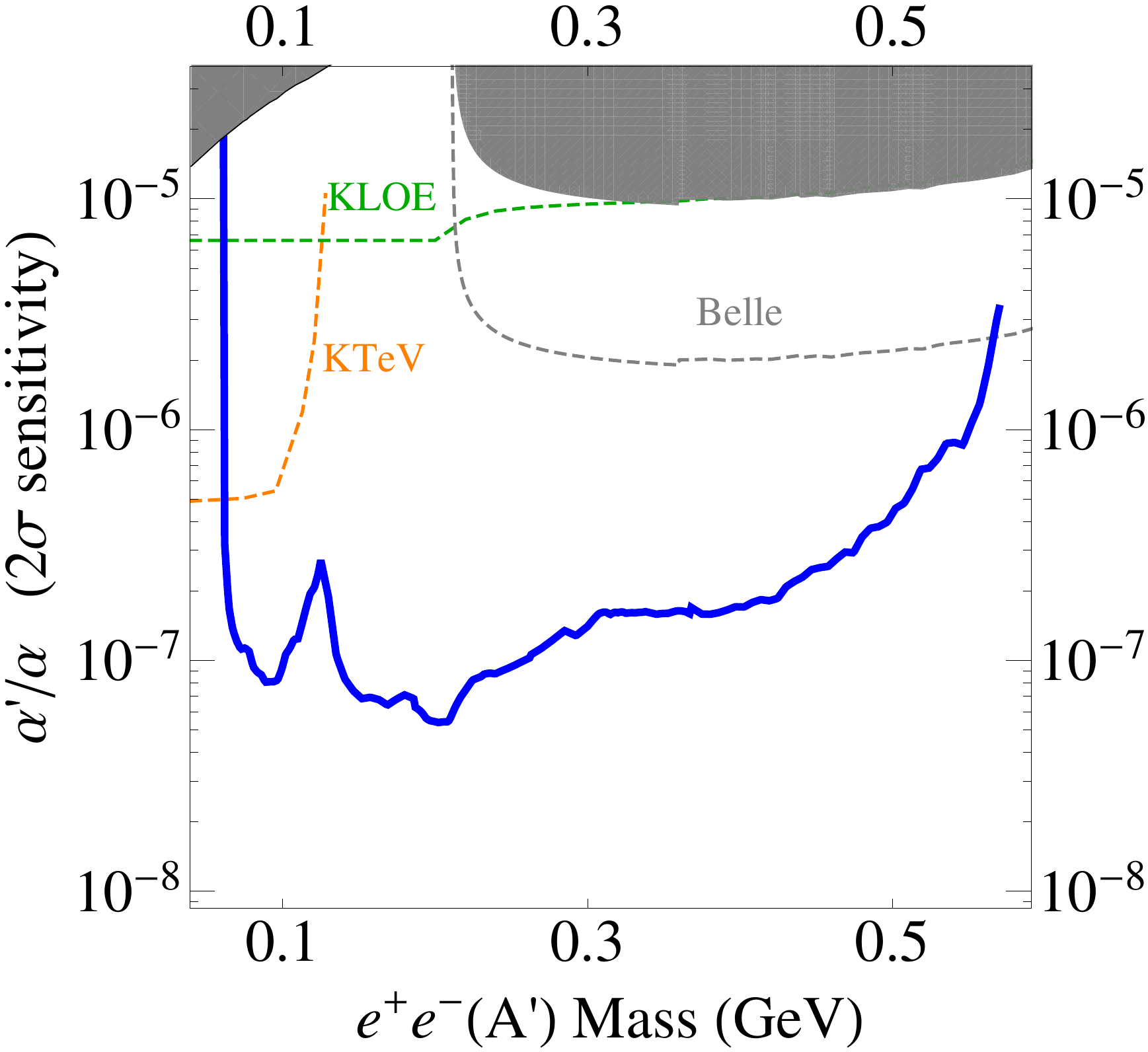}
\end{center}
\caption{\label{fig:potentialSensitivity} Anticipated 2$\sigma$ sensitivity in $\alpha'/\alpha = \epsilon^2$ 
   for the $A'$ experiment (APEX) at Hall A in JLab (thick blue line), compared with current limits
  and estimated
  potential $2\sigma$ sensitivity for $A'$
  searches in existing data (dashed lines), assuming optimal sensitivity as described in the text.  From left to right: KTeV $\pi^0 \rightarrow \gamma A' \rightarrow \gamma
  e^+e^-$ (orange dashed curve), KLOE $\phi \rightarrow \eta A'
  \rightarrow \eta e^+e^-$ (green dashed curve) and Belle $e^+e^-
  \rightarrow \gamma A' \rightarrow \gamma \mu^+\mu^-$ (gray dashed curve).  Existing constraints are as in Figure \ref{fig:bigSummary}.}
\end{figure}

BaBar, BELLE, and KTeV (E799-II) have produced and
detected large numbers of neutral pions, of order $10^{10}$,
of which roughly 1\% decay in the Dalitz mode $\pi^0 \rightarrow e^+e^-
\gamma$.  These experiments can search for the decay $\pi^0
\rightarrow \gamma A'$ induced by $A'$--photon kinetic mixing, which
would appear as a narrow resonance over the continuum Dalitz decay
background.  KTeV has the largest $\pi^0$ sample, and its $e^+e^-$
mass resolution can be approximated from the reported measurement
of the $\pi^0 \rightarrow e^+e^-$ branching fraction \cite{Abouzaid:2006kk}
to be roughly 2 MeV.  This paper also reports the measured mass
distribution of Dalitz decays above 70 MeV, from which we estimate
potential sensitivity to $\alpha'/\alpha$ as small as $5\times 10^{-7}$ for
$70 < m(e^+e^-) \lesssim 100$ MeV, as shown by the orange shaded region 
in Figure \ref{fig:potentialSensitivity}.

Similarly, KLOE can search for the decay $\phi \rightarrow \eta A'$,
likewise induced by $A'$ kinetic mixing with the photon, in a sample
of $10^{10}$ $\phi$'s.  An analysis of this data is ongoing
\cite{Bossi}.  We have taken the blue dashed curve in Figure
\ref{fig:potentialSensitivity} from \cite{Reece:2009un}, which assumes
that mass resolution $\sigma_m$ is dominated by KLOE's $0.4\%$ momentum
resolution.  We have adjusted the contours from \cite{Reece:2009un} to
determine a $2\sigma$ contour and enlarged the bin width used to
determine signal significance from $\sigma_m$ in  \cite{Reece:2009un}
to $2.5\sigma_m$.  Above the muon threshold, $\phi$ decays are not
competitive with $B$-factory continuum production.

In addition, BaBar and Belle can search for the continuum production
mode $e^+e^- \rightarrow \gamma A' \rightarrow \gamma \mu^+\mu^-$ in
their full datasets.  For example, an analysis of the Belle
$\Upsilon(4S)$ data set would increase statistics by a factor of $\sim
24$ relative to the BaBar $\Upsilon(3S)$ search that we have
interpreted as a limit above. 
We have derived the expected sensitivity (shown as a black dashed line
in Figure \ref{fig:potentialSensitivity}) simply by scaling the
$\Upsilon(3S)$ estimated reach by $\sqrt{24}$.  
These searches have not been extended below the muon threshold because
of large conversion backgrounds.

\section{$A'$ production in fixed target interactions}
\label{sec:reaction}

\begin{figure}
\begin{center}
\includegraphics[width=0.375\textwidth]{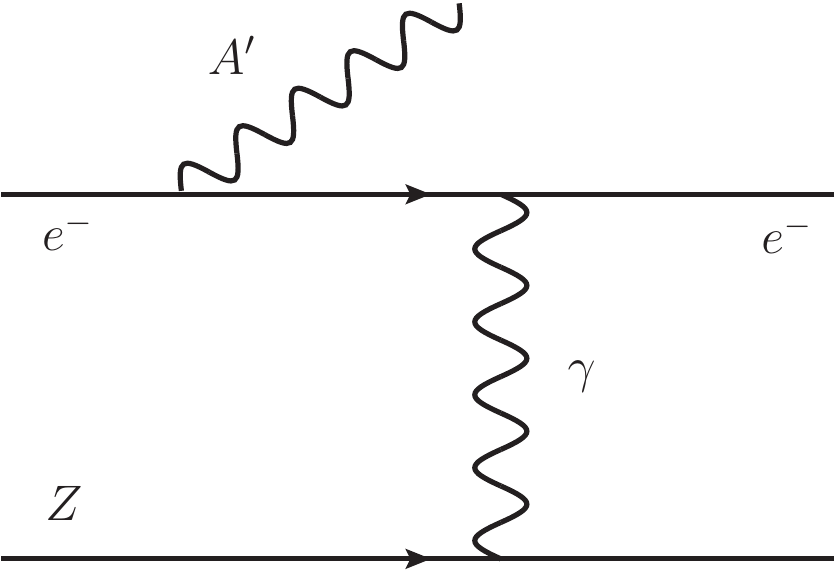}
\end{center}
\caption{$A'$ production by bremsstrahlung off an incoming electron 
scattering off protons in a target with atomic number $Z$.
\label{fig:Sig}}
\end{figure}

$A'$ particles are generated in electron collisions on a fixed target
by a process analogous to ordinary photon bremsstrahlung, see Figure \ref{fig:Sig}. 
This can be reliably estimated in the Weizs\"acker-Williams approximation (see
 \cite{Bjorken:2009mm, Kim:1973he,Tsai:1973py,Tsai:1986tx}).  
When the incoming electron has energy $E_0$, the
differential cross-section to produce an $A'$ of mass $m_{A'}$ with
energy $E_{A'} \equiv x E_0$ is
\bea
\f{d\sigma}{dxd\cos\theta_{A'}} & \approx & \f{8 Z^2\alpha^3 \epsilon^2 E_0^2
  x}{U^2} {\tilde \chi} \times \nonumber\\ 
 & & \!\!\!\!\!\!\!\!\!\!\!\! \!\!\!\!\!\! \!\!\!\!\!\! \!\!\!\!\!\!  \bigg[  (1-x+\f{x^2}{2}) 
 - \f{x (1-x) m_{A'}^2 \left(E_0^2 x\, \theta_{A'}^2\right)}{U^2}
\bigg]\label{eq:dSigmadxdcos}
\eea
where $Z$ is the atomic number of the target atoms, $\alpha \simeq 1/137$, $\theta_{A'}$ 
is the angle in the lab frame between the emitted $A'$ and the incoming electron, \be
U(x,\theta_{A'}) = E_0^2 x \theta_{A'}^2 + m_{A'}^2 \f{1-x}{x} + m_e^2 x \label{Udef}
\ee
is the virtuality of the intermediate electron in initial-state
bremsstrahlung, and $\tilde \chi \sim 0.1-10$ is the Weizs\"acker-Williams effective photon flux, with an overall factor of $Z^2$ removed.    The form of $\tilde\chi$ and its dependence on the $A'$ mass, beam energy, and target nucleus are discussed in Appendix \ref{sec:chi}.  The above results are valid for 
\be
m_e \ll m_{A'} \ll E_0, \qquad  x \,\theta_{A'}^2 \ll 1.
\ee

For $m_{A'} \gg m_e$,  the angular integration gives
\be
\f{d\sigma}{dx} \approx \f{8 Z^2\alpha^3 \epsilon^2 x}{m_{A'}^2} 
\left(1 + \f{x^2}{3 (1-x)}\right) \tilde \chi \label{dSigmadX}.
\ee
The rate and kinematics of $A'$ radiation differ from massless bremsstrahlung in several important ways:
\begin{description}
\item[Rate:] The total $A'$
  production rate is controlled by $\f{\alpha^3 \epsilon^2}{m_{A'}^2}$.
  Therefore, it is suppressed relative to photon bremsstrahlung by
  $\sim \epsilon^2 \f{m_e^2}{m_{A'}^2}$.  Additional suppression from small $\tilde\chi$  occurs for large $m_{A'}$ or small $E_0$.
\item[Angle:] $A'$ emission is dominated at angles $\theta_{A'}$ such that
  $U(x,\theta_{A'}) \lesssim 2 \,U(x,0)$ (beyond this point, wide-angle
  emission falls as $1/\theta_{A'}^4$).  For $x$ near its median value, the
  cutoff emission angle is
\be
\theta_{\rm A'\, max} \sim \max\left(\f{\sqrt{m_{A'} m_e}}{E_0},\f{m_{A'}^{3/2}}{E_0^{3/2}}\right),
\ee
which is parametrically smaller than the opening angle of the 
$A'$ decay products, $\sim m_{A'}/E_0$.  
Although this opening angle is small, the backgrounds mimicking the signal 
(discussed in \S \ref{sec:backgrounds}) dominate at even smaller angles.
\item[Energy:] $A'$ bremsstrahlung is sharply peaked at $x \approx 1$,
  where $U(x,0)$ is minimized.  When an $A'$ is produced, it carries
  nearly the entire beam energy --- in fact the median value of
  $(1-x)$ is $\sim \max\left(\f{m_e}{m_{A'}},\f{m_{A'}}{E_0}\right)$.  
\end{description} 
The latter two properties are quite important in improving signal significance, and are discussed further in \S \ref{sec:backgrounds}.  

Assuming the $A'$ decays into Standard Model particles rather than exotics, its boosted lifetime is 
\bea
\ell_0 & \equiv & \gamma c\tau \simeq \f{3 E_{A'}}{N_{\rm eff} m_{A'}^2 \alpha \epsilon^2} \nonumber \\ 
& \simeq & \f{0.8\mbox{cm}}{N_{\rm eff}} \left ( \f{E_0}{10 \mbox{GeV}} \right ) \!\!
\left (\f{10^{-4}}{\epsilon} \right )^2 \!\!
\left ( \f{100\, \mbox{MeV}}{m_{A'}} \right )^2,
\label{gammaCTau}
\eea
where we have neglected phase-space corrections, and 
$N_{\rm eff}$ counts the number of available decay products.  
If the $A'$ couples only to electrons, $N_{\rm eff} = 1$.  If 
the $A'$ mixes kinetically with the photon, then $N_{\rm eff} = 1$ for $m_{A'}  < 2
m_{\mu}$ when only $A'\to e^+e^-$ decays are possible, and $2+R(m_{A'})$ 
for $m_{A'}\ge2 m_{\mu}$, where 
$R=\f{\sigma(e^+e^- \rightarrow \mbox{ hadrons};\, E=m_{A'})}{\sigma(e^+e^-  \rightarrow \mu^+\mu^-;\, E=m_{A'})}$ 
\cite{Amsler:2008zzb}.    
For the ranges of $\epsilon$ and $m_{A'}$ probed by this experiment,
the mean decay length $\ell_0 \lesssim 250 \mu\rm{m}$ is not significant,
but the ability to cleanly reconstruct vertices displaced forward by a
few cm would open up sensitivity to considerably lower values of
$\epsilon$.

The total number of $A'$ produced when $N_e$ electrons scatter in a
target of $T\ll 1$ radiation lengths is
\be
N \sim N_e \,\f{N_0 X_0 }{A} \,T \, \f{Z^2\alpha^3 \epsilon^2}{m_{A'}^2}\,
\tilde\chi \sim  N_e\, \mathcal{C} \, T\, \epsilon^2 \,\f{m_e^2}{m_{A'}^2},\label{rateApproxThin}
\ee
where $X_0$ is the radiation length of the target in g/cm$^2$, 
$N_0 \simeq 6 \times 10^{23} \,{\rm mole}^{-1}$ is Avogadro's number, and $A$ is the 
target atomic mass in g/mole. The numerical factor $\mathcal{C} \approx 5$ is logarithmically dependent on the
choice of nucleus (at least in the range of masses where the form-factor is
only slowly varying) and on $m_{A'}$, because, roughly, $X_0\propto
\f{A}{Z^2}$ (see \cite{Bjorken:2009mm} and \cite{Amsler:2008zzb}).
For a  Coulomb of incident electrons, the total number of $A'$'s produced is given by 
\be
\f{N}{\mbox{C}} \sim 10^{6} \tilde \chi \left ( \f{T}{0.1} \right ) \left (\f{\epsilon}{10^{-4}} \right)^2 
\left( \f{100\, \mbox{MeV}}{m_{A'}} \right)^2.
\ee

The spectrometer efficiency can be estimated from Monte Carlo simulation of the signal, discussed in \S \ref{sec:signalTrident}.  
It is quite low in APEX, but of course depends on the precise spectrometer settings.  For example, for $m_{A'} = 200$ MeV, $E_0 = 3.056$ GeV, 
an angular acceptance window of $\theta_x = 0.055 - 0.102$ rad and $|\theta_y| \le 0.047$ rad (corresponding 
to an HRS central angle of $4.5^\circ$) and a momentum acceptance of $E = 1.452 - 1.573$ GeV for both the positron and one of the 
electrons, gives a spectrometer efficiency of $\sim 0.14\%$.  

\section{Experimental setup}
\label{sec:exp_setup}

In this section, we describe the experimental setup of the APEX experiment in JLab Hall A.  Many of these features are 
also readily adaptable to other experimental facilities.  

\begin{figure}[t]
\centerline{
\includegraphics[width=0.3\textwidth, angle = 270.]{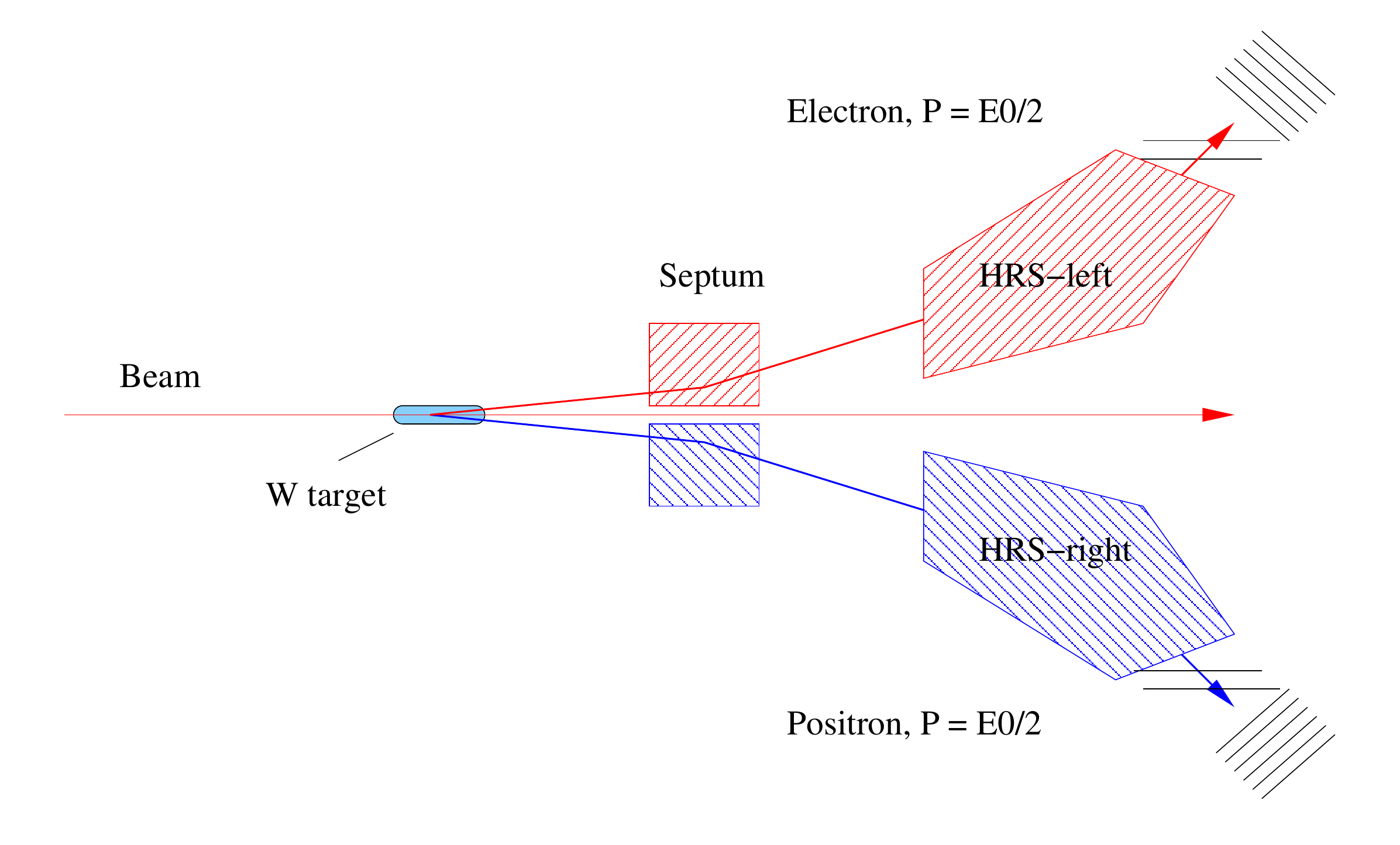}}
\caption{{The layout of the experimental setup --- see text for details.}}
\label{fig:expsetup}
\end{figure}

The APEX experiment will measure the invariant mass spectrum of
$e^+e^-$ pairs produced by an incident beam of electrons on a tungsten
target.
The experiment uses the two high-resolution spectrometers (HRS)~\cite{NIM} 
available in Hall~A at JLab (see Table \ref{tab:spectrospecs} for design specifications), 
together with a septum magnet constructed for the 
PREX experiment~\cite{PREX}, see Figure \ref{fig:expsetup}.  
The physical angle of the HRS with respect to the beam line does not go 
below $\sim 12^\circ$, but the septum allows smaller angles to be probed 
down to $\sim 4^\circ-5^\circ$ by bending charged tracks outward.  
The detector package in each HRS available in JLab Hall A 
includes two vertical drift chambers (VDC),
the single photo-multiplier tube (PMT) trigger scintillator counter (``S0 counter''), 
the Gas Cherenkov counter, the segmented high-resolution 
scintilator hodoscope, and the double-layer lead-glass shower counter.

The electron beam has a current of 80 $\mu A$ (corresponding to $\sim 7$ C on target per day!), 
and will be incident on a solid target located on a target ladder
in a standard scattering chamber.
The target will be made of tungsten wires strung together in
a horizontal plane orthogonal to the beam direction.
The target plane will be mounted at an angle of about 10~mrad with
respect to the horizontal plane.
The beam will be rastered by $\pm$0.25~mm in the horizontal and
$\pm$2.5~mm in the vertical direction to avoid melting the target.

The electron will be detected in the the right HRS (HRS-R) and the positron
will be detected in the left HRS (HRS-L).
The trigger will be formed by a coincidence of two signals from
the S0 counters of the two arms and a coincidence of the signal in the S0 counters 
with a signal from the Gas Cherenkov counter
of the HRS-L (positive polarity arm).
A timing window of 20~ns will be used for the first coincidence and
40~ns for the second coincidence.
The resulting signal will be used as a primary trigger of data acquisition (DAQ).
An additional logic will be arranged with a 100~ns wide coincidence
window between signals from the S0 counters.
This second type of trigger will be prescaled by a factor 20 for DAQ, and is used to 
evaluate the performance of the primary trigger.
Most of the DAQ rate will come from events with a coincident 
electron and positron within a 20~ns time interval.

Note that since we want to search for a narrow peak in the invariant mass spectrum 
of $e^+e^-$ pairs, which requires a high level of statistical precision, it is especially 
important to have a very small level of systematics and a smooth invariant 
mass acceptance.  
In \cite{proposal}, we show that APEX has these properties.  

\begin{figure}[t]
\centerline{
\includegraphics[width=0.2\textwidth, angle = 270.]{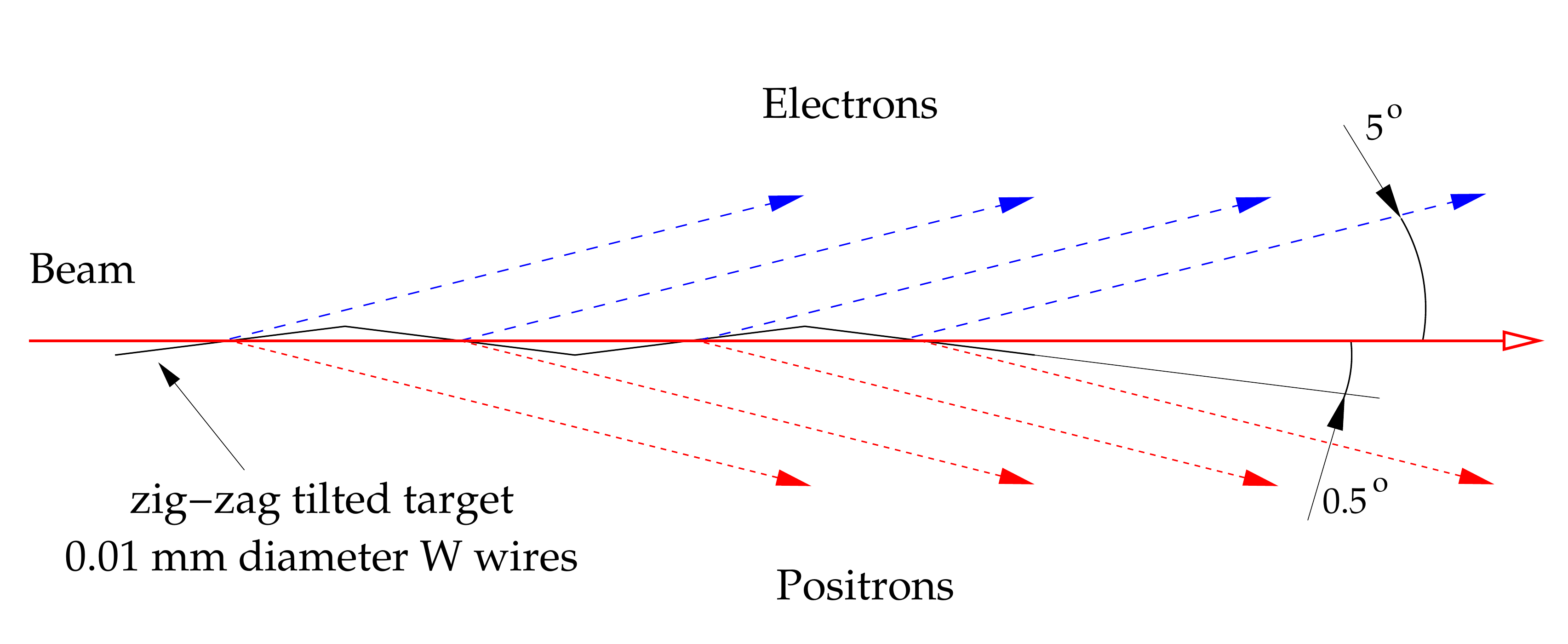}}
\caption{The top view of the tilted target.
The beam is rastered over an area 0.5$\times$5~mm$^2$ 
(the latter is in the vertical direction). 
The beam intersects the target in four areas spread over almost 500~mm.
Pair components will be detected by two HRS spectrometers at a central
angle of $\pm$5$^\circ$. 
Each zig-zag of the target plane is tilted with respect to the beam
by 0.5$^\circ$ and consists of a plane of parallel wires perpendicular to the beam.  
This reduces the multiple scattering of the outgoing $e^+e^-$ pair 
(produced in a prompt $A'$ decay), as described in the text.}
\label{fig:target}
\end{figure}
\begin{figure}[t]
\centerline{
\includegraphics[width=0.45\textwidth, angle = 0]{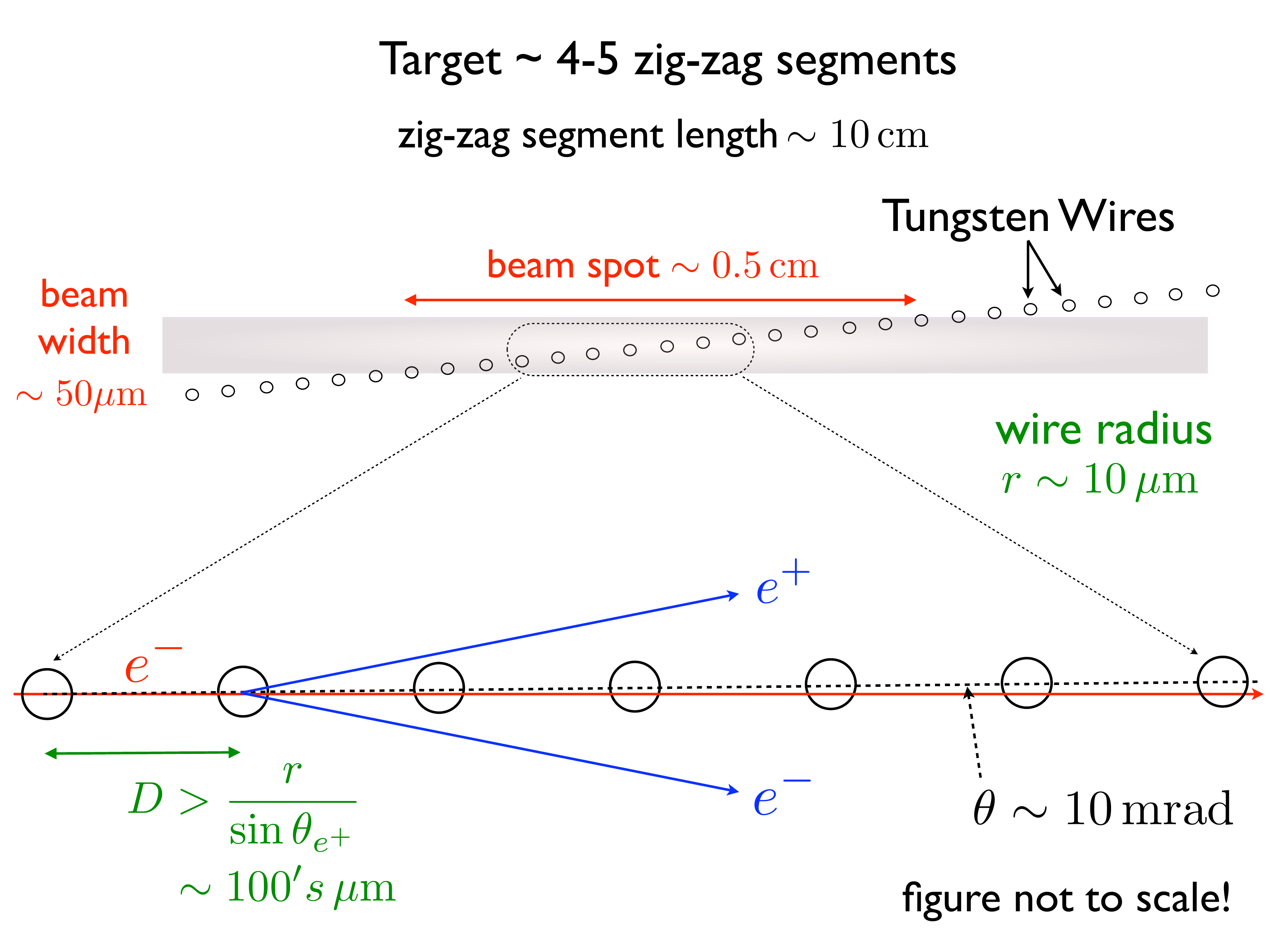}}
\caption{A schematic close-up view of the target.  The figure is not to scale.  The target consists of 4--5 zig-zag planes, with each plane consisting of tungsten wires strung together.   Each zig-zag plane is $\sim 10$ cm long, and lies at an angle of $\sim 10$ mrad with respect to the beam line.  The tungsten wires have a radius of $\sim 10\mu$m and are spaced at a distance of $\sim 100$'s $\mu$m.  While each beam electron can traverse up to $\sim 10$ wires, the production and prompt decay of an $A'$ in a wire produces $e^+e^-$ pairs that have an angle of $\sim m_{A'}/E_{\rm beam}$, large enough for them to miss the next wire --- this greatly reduces the multiple scattering, and is the reason for not using a target foil.  The beam width is $\sim 50 \mu$m, which translates into a $\sim 0.5$cm large beam spot along the target plane.  The vertical rastering of the beam of $\sim 0.5$mm moves the beam spot $\sim 5$cm back-and-forth along the target plane --- this helps to prevent the beam from melting the target. }
\label{fig:target2}
\end{figure}
%

\subsection{The long tilted target}

\begin{table}[ht]
\caption{Main design characteristics of the Hall~A 
High Resolution Spectrometers at nominal target position 
(see \cite{NIM} for more details).  
The resolution values are for the full-width at half-maximum. 
These parameters correspond to a point target and do not include 
the effects of multiple scattering in the target and windows.
In the calculation of the invariant mass resolution the effect of 
multiple scattering in the target was taken into account.
The vacuum coupling of the scattering chamber and the spectrometer
allows one to avoid using windows.
}
\vskip 0.25cm
\label{tab:spectrospecs}
\begin{center}
\begin{tabular}{|l|c|}
  \hline
Configuration	& QQD$_{n}$Q Vertical bend\\
Bending angle	&45$^{\circ}$\\
Optical length	&23.4 m\\
Momentum range	&0.3 - 4.0 GeV/c\\
Momentum acceptance	&-4.5$\% < \delta$p/p $<$+4.5\%\\
Momentum resolution	& 1$\times$10$^{-4}$\\
Dispersion at the focus (D)  &12.4 m\\
Radial linear magnification (M)	&-2.5\\
           D/M	&5.0\\
Angular range HRS-L	&12.5$^{\circ}$ - 150$^{\circ}$\\
\hspace{2.0cm} HRS-R	&12.5$^{\circ}$ - 130$^{\circ}$\\
Angular acceptance: {\hfill Horizontal}	&$\pm$30 mrad\\
{\hfill Vertical}	                &$\pm$60 mrad\\
Angular resolution :{\hfill Horizontal}	&0.5 mrad\\
{\hfill Vertical}	                &1.0 mrad\\
Solid angle at $\delta$p/p = 0, y$_{0}$ = 0	&6 msr\\
Transverse length acceptance	&$\pm$5 cm\\
Transverse position resolution	&1 mm\\
\hline
\end{tabular}
\end{center}
\end{table}

\ind The experiment will utilize the standard Hall A scattering
chamber as it is used by the PREX experiment, with a target consisting 
of a 50-cm-long tilted wire mesh plane.
The concept of the target is presented in Figures \ref{fig:target} and \ref{fig:target2}.
The wires comprising each plane are perpendicular to the beam-line. 
The tilt angle of 10~mrad is sufficient to ensure stability of 
the beam-target geometry, and at the same time such a tilt angle 
is 10 times smaller than the central angle to the HRS, 
which results in a reduction of the path length traversed by 
the produced $e^+e^-$ pairs. 
The wires comprising of each zig-zag plane are spaced so that outgoing $e^+e^-$ pairs coming from prompt 
$A'$ decays inside a wire only travel through a single wire (for some configurations, the outgoing $e^+e^-$ 
pair may not have to traverse any wire if the $A'$ does not decay inside a wire).  
For wire thickness of $\sim 10^{-3}$ radiation lengths, 
this considerably reduces the multiple scattering in the target versus that in a true foil and 
leads to a much better mass resolution. The maximum number of wires that a beam electron can pass 
through per plane is $\sim 10$ in the configuration illustrated in Figure \ref{fig:target} or Figure \ref{fig:target2} 
assuming $10 \mu$m diameter
tungsten wires. Wires as thick as $15 \mu$m can be used without significantly compromising mass resolution. 

The plane of the target wire mesh will be vertical.
The mesh plane will have 4--5 zig-zags, each with length $5-10$ cm, which result
in multiple intersection points allowing an extra factor of 4--5 for 
rejection of accidental tracks in offline analysis.  (Figure \ref{fig:target} illustrates a target with 4 
zig-zags.)

The central angle of the spectrometer varies with the position 
of the target.
In this experiment, such a variation is very useful because 
it extends the range of invariant mass covered with one setting 
of the spectrometers. For several settings suggested in our run 
plan, only two planes of wire mesh are needed, 
one at the front and one at the back of the acceptance region. 

There are two considerations to take into account when selecting the
material.  
The first consideration is to achieve the highest possible
ratio of signal to background, while keeping the background rate low
enough so as not to overwhelm the triggering and DAQ  
system.  
The second is whether a thin foil or a thin wire of a
particular material is available.  Large backgrounds come from pions
produced in photo-production from nucleons, and from electrons
produced in the radiative tail of electron-proton elastic scattering.
These backgrounds do not mimic the signal, but if their rate is too
large, they can overwhelm the DAQ system.  
These considerations favor
the use of a tungsten target, with a total thickness between 0.5\%
and 10\% radiation length, with thicker targets used in higher-energy
runs.  Reduction of the thickness at low energies is required to limit
the rate in the electron spectrometer and also minimizes the
multiple-scattering contribution to the pair mass resolution.  

The heat load of the target is also an important consideration.  
This is mitigated by rastering the beam and using 
materials like tantalum or tungsten.
The tilted target cools from a large area; for example, 
with the raster size $0.5\times 5$~mm and proposed geometry 
(Figure~\ref{fig:target}), the cooling area is 20~cm$^2$.
For the parameters of APEX (an electron beam of 80~$\mu$A 
on a 10\%~$X_0$ tungsten target) the head load is about 140~W
(or 7~W/cm$^2$), 
which results in the equilibrium target temperature of 1000$^\circ$K.
Experimental study has demonstrated that 1~kW/cm$^2$ is a safe level 
for a tungsten foil target~\cite{Perez}, so we expect that the wire mesh target 
will perform quite well. 
\section{Signal and Trident Kinematics}
\label{sec:signalTrident}

The stark kinematic differences between QED trident backgrounds and the $A'$ signal are the 
primary considerations in determining the momentum settings of the spectrometers.  
As we will show in \S \ref{sec:backgrounds}, QED tridents dominate the final 
event sample after offline rejection of accidentals, so we consider their properties in some detail here. 

\begin{figure}[t]
\begin{center}
\includegraphics[width=0.47\textwidth]{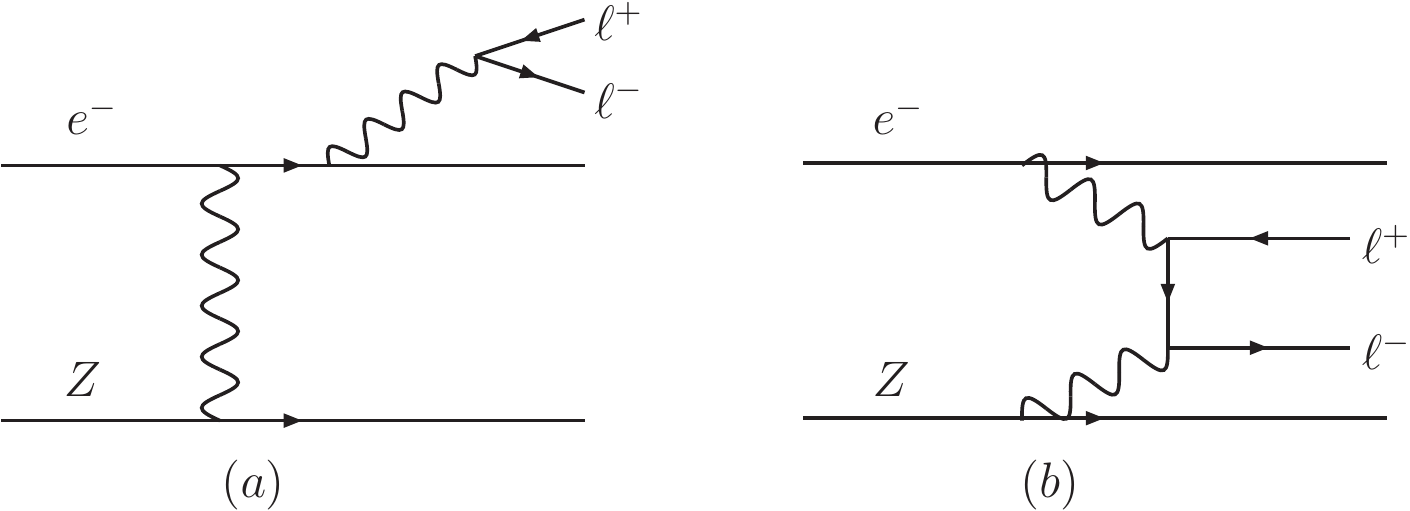}
\end{center}
\caption{Sample diagrams of (a) radiative trident ($\gamma^*$) and (b) Bethe-Heitler 
trident reactions that comprise the primary QED background to $A'\rightarrow \ell^+\ell^-$ search channels.
\label{fig:diagrams}}
\end{figure}

The irreducible background rates are given by the diagrams shown in Figure \ref{fig:diagrams}.  
These trident events can be usefully separated into ``radiative'' diagrams (Figure \ref{fig:diagrams}(a)),
 and ``Bethe-Heitler'' diagrams (Figure \ref{fig:diagrams}(b)), that are separately gauge-invariant.  

We have simulated the production of these continuum trident background events in QED using the nuclear 
elastic and inelastic form-factors in \cite{Kim:1973he}.  The simulation is done using 
MadGraph and MadEvent~\cite{Alwall:2007st} to compute the matrix elements for 
$e^- Z \rightarrow e^-\ ( e^+ e^-)\ Z$ exactly, but neglecting the effect of nuclear 
excitations on the kinematics  in inelastic processes.  The MadEvent code was 
modified to properly account for the masses of the incoming nucleus and 
electron in event kinematics, and the nucleus is assumed to couple with a 
form-factor $G_2$ defined in Appendix A.  

The continuum trident background was simulated including the full interference effects between the diagrams in 
Figure \ref{fig:diagrams}.  
In addition, a ``reduced-interference'' approximation simplifies the analysis and is much less computationally 
intensive.  
In this approximation, we treat the recoiling $e^-$ and the $e^-$ from the produced pair as distinguishable.  
Furthermore, we separate trident processes into the radiative diagrams (Figure \ref{fig:diagrams}(a)) and 
 the Bethe-Heitler diagrams (Figure \ref{fig:diagrams}(b)), and 
 we calculate the cross-section for both of these diagrams separately.  
This approximation under-estimates the background rates by a factor of about 2--3 in the range of $A'$ 
masses and beam energies considered in this paper and \cite{proposal}.  
For the reach analysis discussed below, we have used 
differential distributions computed  in the ``reduced-interference'' approximation, then rescaled to the 
cross-section for the full-interference process.

The contribution from the radiative diagrams (Figure \ref{fig:diagrams}(a)) alone is also useful 
as a guide to the behavior of $A'$ signals at various masses.  Indeed, the kinematics of the $A'$ 
signal events is identical to the distribution of radiative trident events restricted in an invariant mass 
window near the $A'$ mass.  Moreover, the rate of the $A'$ signal is simply related to the radiative trident 
cross-section within the spectrometer acceptance and a mass window of width $\delta m$ by \cite{Bjorken:2009mm}
\be
\f{d\sigma(e^-Z\rightarrow e^- Z (A'\to \ell^+\ell^-))}{d\sigma(e^-Z\rightarrow e^- Z (\gamma^*\to \ell^+\ell^-))}
=  \left ( \f{3\pi\epsilon^2}{2N_{\rm eff}\alpha} \right ) \left ( \f{m_{A'}}{\delta m} \right), \label{eq:theorem}
\ee
where $N_{\rm eff}$ counts the number of available decay products and is defined below equation (\ref{gammaCTau}).
This exact analytic formula was also checked with a MC simulation of both the $A'$ signal and the radiative tridents background 
restricted to a small mass window $\delta m$, and we find nearly perfect agreement.  
Thus, the radiative subsample can be used to analyze the signal, which simplifies the analysis considerably.  

It is instructive to compare kinematic features of the radiative and
Bethe-Heitler distributions, as the most sensitive experiment
maximizes acceptance of radiative events and rejection of
Bethe-Heitler tridents.  Although the Bethe-Heitler process has a much
larger total cross-sections than either the signal or the radiative
trident background, it can be significantly reduced by exploiting its
very different kinematics.  In particular, the $A'$ carries most of
the beam energy (see discussion in \S \ref{sec:reaction}), while the
recoiling electron is very soft and scatters to a wide angle.  In
contrast, the Bethe-Heitler process is not enhanced at high pair
energies.  Moreover, Bethe-Heitler processes have a forward
singularity that strongly favors asymmetric configurations with one
energetic, forward electron or positron and the other constituent of
the pair much softer.

\begin{figure}[!t]
\begin{center}
\includegraphics[width=0.45\textwidth]{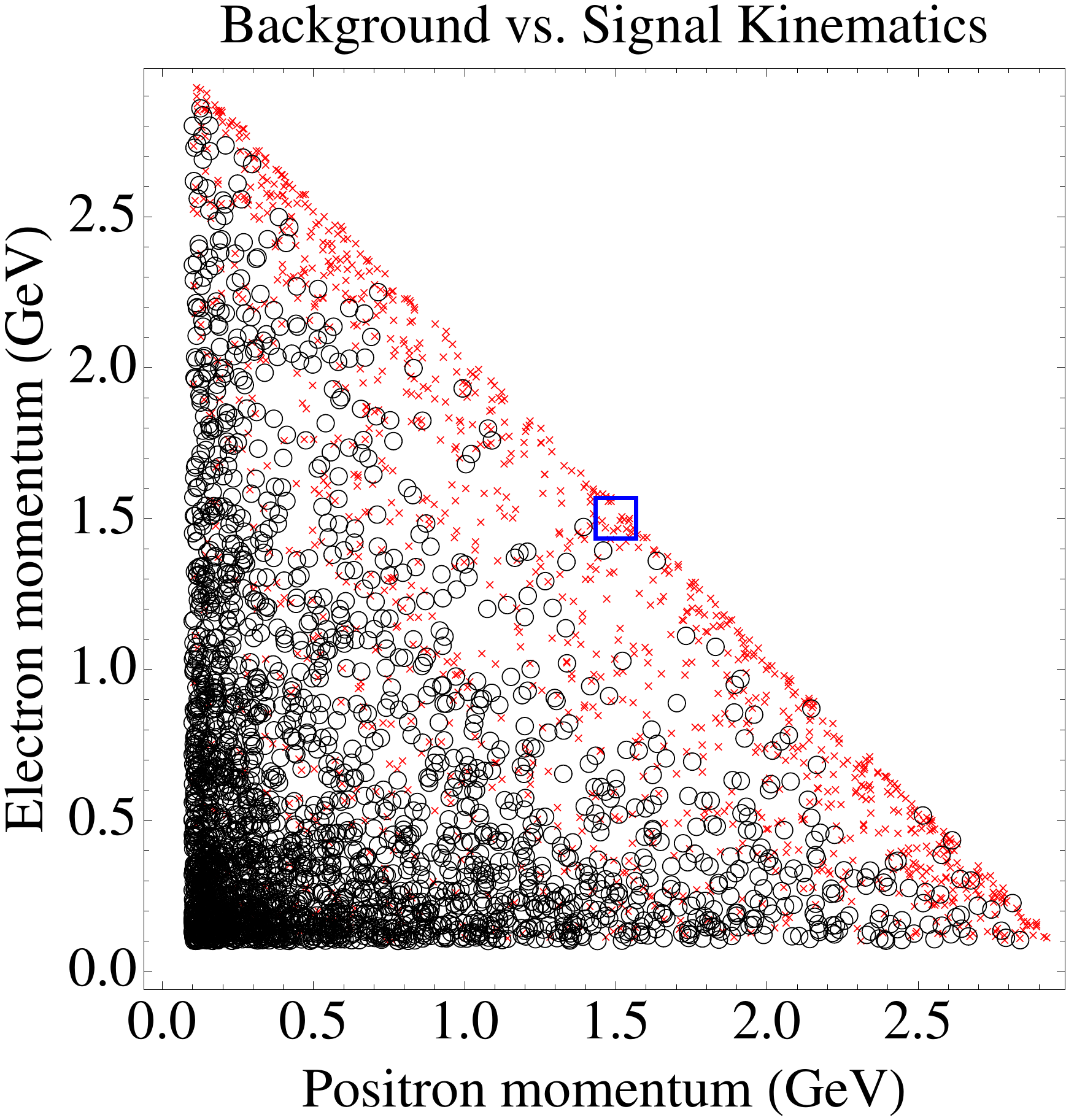}
\end{center}
\caption{
\label{fig:momenta}
Positron and electron momenta in $A'$ signal events with $m_{A'}=200$ MeV (red crosses) and in Bethe-Heitler background events, for a 3 GeV beam energy.  Comparably sized signal and Bethe-Heitler samples were used to highlight the kinematics of both; in fact the expected signals are much lower than the Bethe-Heitler process 
(see Figure \ref{fig:bumpAndBkg}).  The clustering of $A'$ events at high momenta near the kinematic limit and of Bethe-Heitler events along both axes are evident. A spectrometer acceptance window that optimizes signal sensitivity is indicated by the blue box.
}
\end{figure}

These properties are discussed further in the Appendix of
\cite{Bjorken:2009mm}, and illustrated in Figure \ref{fig:momenta}, which shows a
scatterplot of the energy of the positron and the higher-energy
electron for the signal (red crosses) and Bethe-Heitler background
(black dots).  The signal electron-positron pairs are clearly concentrated
near the kinematic limit, $E(e^+)+E(e^-) \approx E_{beam}$.
Background rejection is optimized in symmetric configurations with
equal angles for the two spectrometers and momentum acceptance of each
spectrometer close to half the beam energy (blue box).

While the signal over background ($S/B$) can be significantly improved with a judicious choice of 
kinematic cuts, the final $S/B$ in a small resolution limited mass window is still very low, 
$\sim 1\%$.  A ``bump-hunt'' for a small 
signal peak over the continuous background needs to be performed. 
This requires 
an excellent mass resolution, which has an important impact on target design and calls for a target that is tilted with 
respect to the beam line (see Appendix B for a discussion of the mass resolution).  

\subsection{Calculation of the $\epsilon$ reach}

\begin{table*}[t!]
\begin{center}
\begin{tabular}{|l|ll|ll|}
\hline
{\bf Settings}                      &	{\bf A}         &	{\bf B}              	&	{\bf C}		&	{\bf D}	               \\
\hline
\sq Beam energy (GeV)			&	2.302		    &	4.482                	&	1.1			&	3.3			         \\
\sq Central angle		   	  	&	5.0\de	    &	5.5\de	 	   	&	5.0\de		&	5.0\de			   \\
\sq Effective angles	   	  	&	(4.5,5.5)	    &	(5.25,6.0)		   	&	(4.5,5.5)		&	(4.5,5.5)			   \\
\sq Target $T/X_0$ (ratio$^{a}$)      	  	&	4.25\% (1:1)    &	10\% (1:1)           	&	0.58\% (1:3)	&	10\% (1:1)		         \\
\sq Beam current (\muA)    	  	&	80		    &	80			   	&	80	     		&	80				   \\
\sq Central momentum (GeV) 	  	&	1.145	          & 2.230				&	0.545			&	1.634				   \\
\hline
{\bf Singles (negative polarity)} &&&& \\ 			                               	 		   		 
\sq $e^-$ (MHz)		   	      &	4.5		    &	0.7			   	&	6.			&	2.9				   \\
\sq $\pi^-$\, (kHz)	   		&	640.		    &	2200			   	&	36.			&	2500.				   \\
\hline			   					                               	                         
{\bf Singles (positive polarity)}  &&&& \\ 			                               	 		   		 
\sq $\pi^+$+$p$\, (kHz)   	      &	640.		    &	2200			   	&	36.			&	2500.				   \\
\sq $e^+$ (kHz) 		   		&	31.		    &	3.6			   	&	24.			&	23.				   \\
\hline 			   					                               	                         
{\bf Trigger/DAQ:}                 &&&& \\	   					                               	                         
\sq Trigger$^{b}$ (kHz)			&	4.		    &	0.4			   	&	3.2			&	3.4				   \\
\hline 			   					                               	                         
{\bf Signal to background:}        &&&& \\	 			                               	   				 
\sq Trident (Hz)		    		&	610		    &	70    		   	&	350			&	530				   \\
\sq Two-step (Hz)  	   		&	35		    &	15			   	&	5			&	75				   \\
\sq Background$^{c}$ (Hz)  		&	70		    &	1.3			   	&	70			&	35				   \\
\hline
\end{tabular}
\vspace{0.1in}
{\small
\begin{tabular}{cp{5.5in}}
$^{a}$ & The listed total target thickness is split between two sets of wire mesh planes, located at different $z$ to produce the two indicated effective angles.  The numbers in parentheses denote the ratio of target thickness at the larger effective angle to that at the smaller effective angle. \\ 
$^{b}$ & Trigger: Coincidence with 20 ns time window between S0-N (assuming pions are rejected by a factor of 100) and S0-P signals. \\
$^{c}$ & Dominated by $e^+ e^-$ accidental rate.  We assume pion rejection by a factor of $10^4$ in offline cuts, a 2 ns time window and additional factor of 4 rejection of accidentals from the target vertex. Further rejection using kinematics is expected, but not included in the table.
\end{tabular}
}
\caption{\label{tab:bigBGtable}
Expected counting rates for the proposed experiment.}
\end{center}
\end{table*}

For all cross sections and rates of reactions described in this paper and \cite{proposal}, 
Monte Carlo based calculations were performed over a grid of 
beam energy settings and central spectrometer angular settings. Interpolation was used to extend this grid continuously to intermediate beam energies and angles --- all rates exhibited expected power law behavior, thereby providing confidence in the reliability of an interpolation. Additional cross checks at specific points were performed to test the accuracy of our interpolation, which was generally better than $\sim 5\%$.

In order to calculate the $\alpha'/\alpha$ reach for a particular choice of 
target nucleus, spectrometer angular setting, profile of wire mesh target, and momentum bite, the following procedure is performed: 
\begin{itemize}
\item Monte Carlo events are simulated for the Bethe-Heitler, radiative tridents, and the 
continuum trident background including the full interference effects between the diagrams.  
The latter background is computationally intensive, and only a small statistics sample is generated, sufficient  
to obtain the cross-section from MadEvent.   
\item The cross-section ratio of the full continuum background (with interference effects) to the sum of the Bethe-Heitler 
and radiative tridents is calculated, and represents a multiplicative factor by which the latter must be multiplied 
to get the background cross-section.  
\item The rates of all reactions impinging the spectrometer acceptance were calculated by integrating over a chosen target profile,
which usually extended from $4.5$ to $5.5$ degrees. For Bethe-Heitler, radiative tridents, and the 
continuum trident background, the calculation of the rate was performed as a function of the invariant mass 
of the $e^+e^-$ pairs. 
\item Using the expressions in Appendix B, we calculated the mass resolution $\delta_m$. 
We then tiled the acceptance region with bins of size $2.5\times \delta_m$ in invariant mass. 
\item As a function of $\alpha'/\alpha$, the total number of signal ($S$) and background ($B$) events was calculated with the help of (\ref{eq:theorem}) for each bin. 
\item We then set $S/\sqrt{B}=2$, and solved for $\alpha'/\alpha$.
\end{itemize} 

This procedure was used to calculate the reach in the $\alpha'/\alpha$ and $m_{A'}$ parameter space 
shown in \S \ref{sec:proposed}.  
Further improvements may be obtained by more sophisticated analysis cuts such as the use of 
matrix element methods (see e.g.~\cite{Freytsis:2009bh}).

\section{Backgrounds}
\label{sec:backgrounds}

In this section, we present the results of an analysis of the expected backgrounds for the $A'$ search.  
Table \ref{tab:bigBGtable} summarizes the expected singles rates, trigger rates, and 
coincidence rates.   
For more details on how we calculated the background rates we refer the reader to \cite{proposal}.

Important backgrounds come from electron, pion, and positron singles.
There are three contributions to the electron singles rate in the HRS at the proposed
momentum settings, namely inelastic scattering, radiative elastic electron-nuclei scattering, and 
radiative quasi-elastic electron-nucleon scattering.
Our calculations of the electron, pion, and positron singles rates were checked 
against measurements made by experiment E03-012 for a 5 GeV electron beam
incident on a hydrogen target, at 6$^\circ$ 2-GeV HRS setting.
The final values of the electron and pion rates were obtained by means
of the ``Wiser'' code~\cite{wiser}; positron singles rates from
trident reactions were calculated using MadGraph and MadEvent~\cite{Alwall:2007st},
described in \S \ref{sec:signalTrident}.  

Using our calculations of the singles rates, we compute the rate of accidental coincident triggers 
arising from an $e^+$ in the HRS-R and an $e^-$ in the HRS-L within the trigger timing 
windows.  
These accidental coincidences are a dominant part of the recorded events in APEX, 
and determine the maximum rate at which potential signal trident events can be recorded.
A typical composition of the single rate in the spectrometers is
expected to be $e^-/\pi^-\approx$~80/20 in the negative polarity arm 
and $\pi^+/p/e^+\approx$~80/19/1 in the positive polarity arm.
The fraction of the true coincidence events could be up to 50\%  for
the $e^-\pi^+$ rate within a 2~ns time window, and could be significant for the 
$e^-p$ events in certain regions of momenta.

Besides the trident events discussed in \S \ref{sec:signalTrident}, an additional 
source of true coincidence events is the ``two-step'' (incoherent) trident process, 
in which an electron radiates a real, hard photon in the target that subsequently 
converts to a high-mass $e^+e^-$ pair.  
For thin targets, this process is suppressed compared to the trident rate, and so it is 
sub-dominant for all the settings we consider.  

The consideration of these rates determine trigger rates and upper bounds on
offline accidental rates shown in Table \ref{tab:bigBGtable}.  

\section{Measurements and Reach in APEX} 
\label{sec:proposed}

\begin{figure}[t]
\begin{center}
\includegraphics[width=0.46\textwidth]{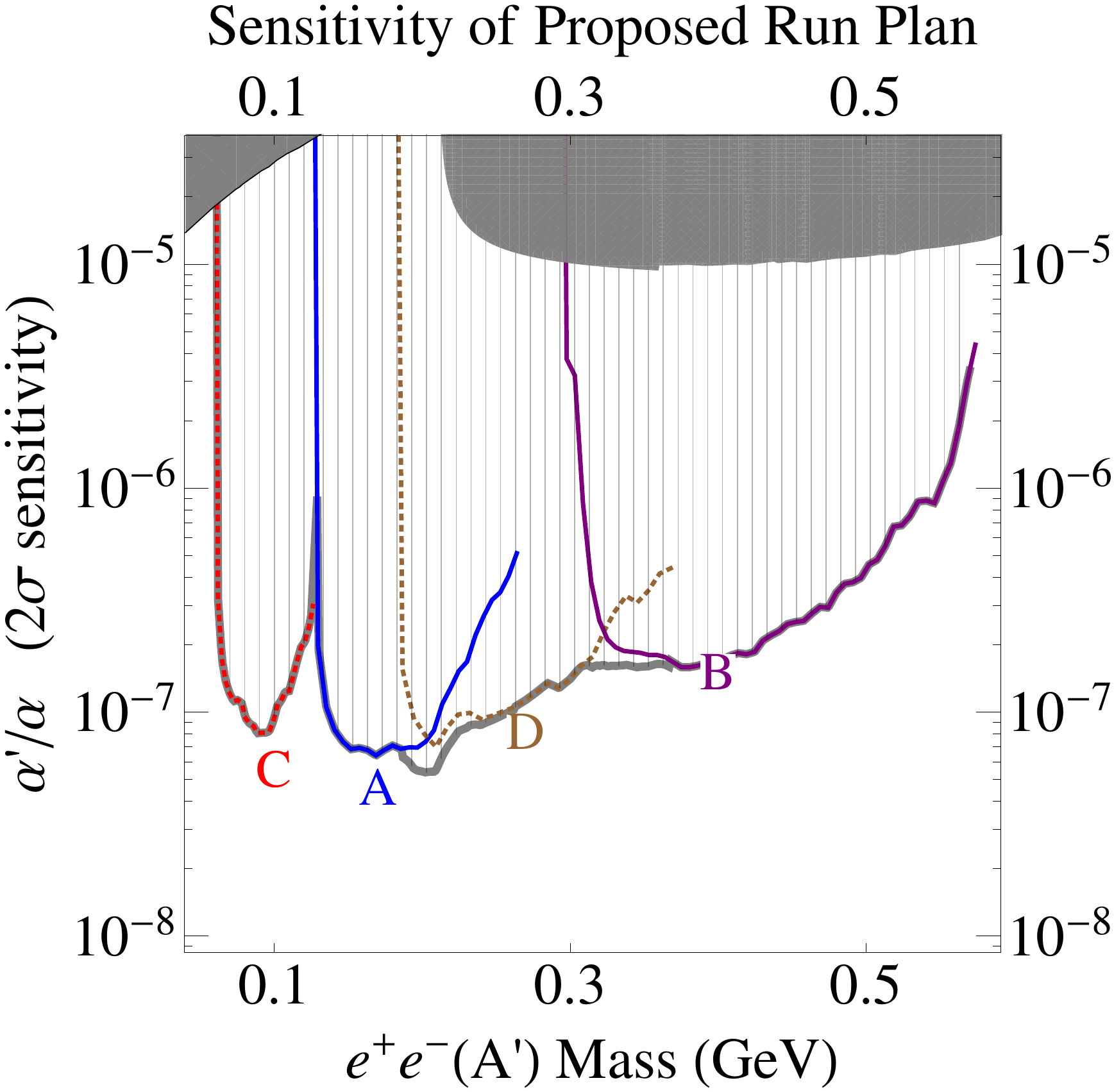}
\caption{Anticipated 2$\sigma$ sensitivity in $\alpha'/\alpha = \epsilon^2$ 
   for APEX \cite{proposal} for the settings given in Table \ref{tab:bigBGtable} 
   (assuming a six-day run in configuration ``A'', ``C'', and ``D'' and a twelve-day run in 
   ``B'').  Existing constraints are shown in the gray shaded regions.   
   The colored curves correspond to the sensitivity in each of the 
   individual energy settings, and the thick gray curve reflects the sensitivity of a combined analysis. 
\label{fig:RunPlan}}
\end{center}
\end{figure}

\begin{figure}[t]
\begin{center}
\includegraphics[width=0.46\textwidth]{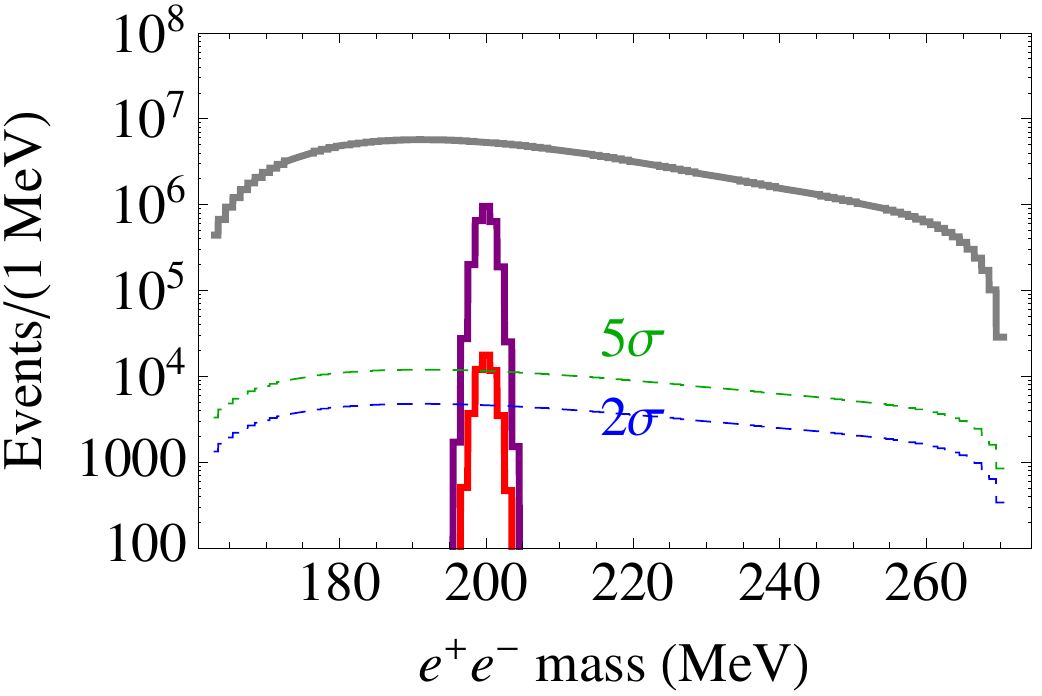}
\includegraphics[width=0.46\textwidth]{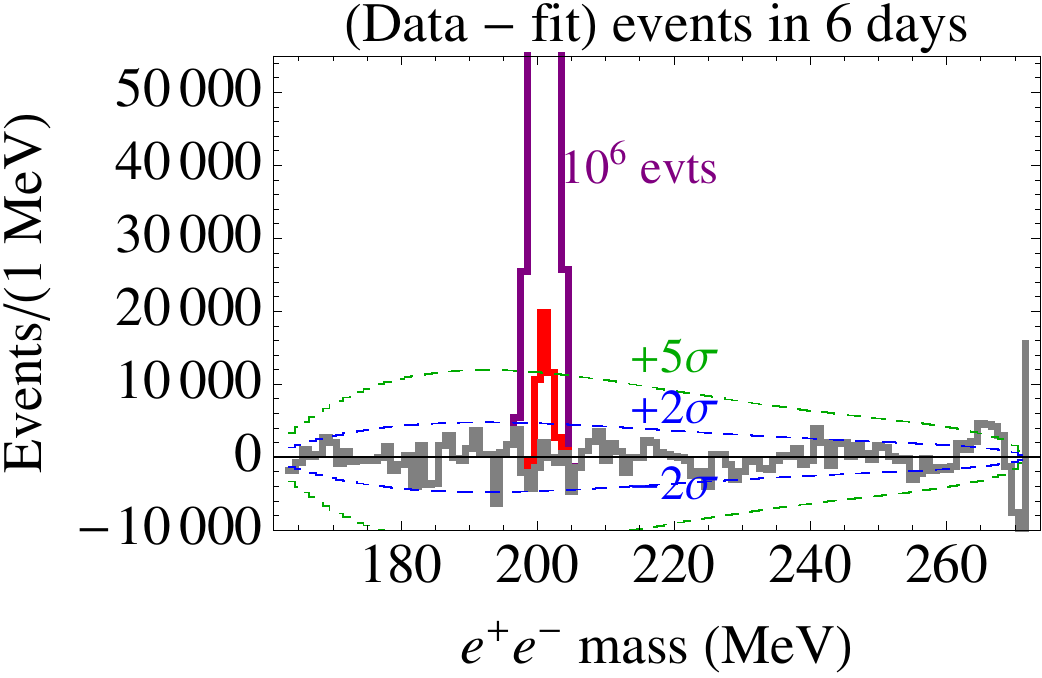}
\caption{Comparison of signal rates in six days of running at 
  setting ``A'' to expected background and statistical
  sensitivity.  {\bf Top:} The resonances in purple and red
  lines correspond to $A'$ signals at 200 MeV, smeared by a Gaussian
  to model detector resolution and multiple scattering, with $\alpha'/\alpha = 6.5\times 10^{-6}$ and $1.3 \times 10^{-7}$, respectively.  The upper
  (purple) signal is just beyond the $2\sigma$ expected sensitivity of a KLOE analysis, while the lower (red) signal corresponds to
  the ``$5\sigma$'' sensitivity (not including a trials factor) of
  this experiment.  The gray line is the simulated invariant mass
  distribution for the continuum trident background, and the blue and
  green dashed lines reflect the size of 2 and $5\sigma$ Poisson
  fluctuations.  {\bf Bottom:} The gray line corresponds to the
  bin-by-bin differences between pseudodata containing no signal and a
  smooth fit to this pseudodata.  Analogous subtractions when a signal
  is present are shown in purple and red, with the same $\alpha'/\alpha$ as
  in the top figure.  Again the blue and green dashed lines reflect
  the size of 2 and $5\sigma$ Poisson fluctuations.
\label{fig:bumpAndBkg}}
\end{center}
\end{figure}

We consider a twelve-day run in the configuration ``B'' of Table 
\ref{tab:bigBGtable} and six-day
runs in each of the remaining configurations, to search for new
resonances in $e^+e^-$ trident spectra from 65 to 550 MeV.  
For settings ``A'' and ``C'', the target thickness and beam current have
been optimized to accumulate the largest possible sample of trident
events without saturating the data acquisition system.  Settings ``B''
and ``D'' are far from data acquisition limits, but we do not use
$T/X_0>10\%$ to avoid limits on the total radiation produced (this can 
possibly be side-stepped at other facilities).

The mass range from 65 to 550 MeV is chosen to take advantage of the
Hall A HRS spectrometers, as well as for its theoretical interest.
Lower masses are more effectively probed by using lower beam energies, 
improved forward acceptance, and/or vertexing  
(see also \cite{Freytsis:2009bh}).
Settings at higher masses are possible but have significantly reduced
sensitivity and are better suited to exploration with
higher-acceptance equipment and an experiment optimized to accept muon
and pion pairs as well as electrons.

In each setting, the proposed experiment will accumulate between 70
and 300 million trident events.  With these statistics, it will be
possible to search offline for small resonances comprising a few
thousandths of the collected data in a resolution-limited window.
This will allow sensitivity to new gauge boson couplings $\alpha'/\alpha$ as low as
$10^{-7}$ over the broad mass range probed by APEX, as
summarized in Figure \ref{fig:RunPlan}.
This sensitivity would improve on the cross-section limits from past experiments by a factor of $\sim 10-1000$.

As a specific example, we have illustrated the expected sensitivity of
setting ``A'' to $A'$ signals with different $\epsilon$ in
Figure \ref{fig:bumpAndBkg}.  Each component of the target populates a different invariant mass distribution; for simplicity 
we consider only the contribution from the front planes of the target, with $\theta_{eff} \approx 5.5^\circ$ 
(recall that the target is extended along the beam line and consists of 4--5 planes in a zig-zag configuration).  
The top panel 
illustrates the absolute
size of $A'$ signals at $m_{A'}=200$ MeV compared to the continuum
trident background (gray line) and the size of $2$ and $5$-sigma
statistical fluctuations (blue and green dashed lines), while the
bottom panel illustrates how the same signals would appear after
subtracting a smooth parameterization of the background.  The purple
curves in each panel corresponds to an $A'$ signal with $\alpha'/\alpha = 7\times10^{-6}$ at 200 MeV, which according to the estimates in \S \ref{ssec:limits} would not be seen or excluded at $2\sigma$ by a future KLOE search in $\phi
\rightarrow \eta A'$.  The red curve has $\alpha'/\alpha = 1.3 \times 10^{-7}$, corresponding to the expected 
``$5\sigma$'' sensitivity (not accounting for the trials factor) in
APEX. 

\section{Conclusions}
\label{sec:conclusion}

This paper summarized a new experiment (``$A'$ \emph{experiment}'', or ``APEX'') that has 
been proposed to the Jefferson Laboratory's PAC 35 \cite{proposal}. 
The experiment proposes to use 30 days of beam 
to measure the electron-positron pair mass 
spectrum and search for new gauge bosons $A'$ in the mass range 65 MeV $<m_{A'} < $ 550 MeV 
that have weak coupling to the electron.  
Parametrizing this coupling by the ratio $\alpha'/\alpha$ that controls the $A'$ production cross-section, 
this experiment would probe $\alpha'/\alpha$ as small as $\sim (6-8)\times 10^{-8}$ at 
masses from 65 to 300 MeV, and $\alpha'/\alpha\sim (2-3)\times 10^{-7}$ at masses up to 525 MeV, making it 
sensitive to production rates 10--1000 
times lower than the best current limits set by measurements 
of the anomalous muon magnetic moment and by direct searches at BaBar.  
The experiment uses the JLab electron beam in Hall A at energies of
about 1, 2, 3, and 4 GeV incident on a long (50 cm) thin tilted tungsten wire mesh target, 
and both arms of the High Resolution Spectrometer at angles between $5.0^\circ$ and $5.5^\circ$ 
relative to the nominal target position. 
The experiment can determine the mass of an $A'$ to an accuracy of $\sim$ 1--2 MeV. 

While this paper was motivated by a specific experimental proposal for JLab Hall A, very similar 
experiments are possible at other experimental facilities, such as the Mainz Microtron or JLab Hall B.  
Many of the considerations discussed in this paper are applicable to these other facilities.

Constraints on new vector bosons with mass near 50 MeV -- 1 GeV are
remarkably weak.  However, such light force carriers are well motivated
theoretically, and several recent anomalies from terrestrial and
satellite experiments suggest that dark matter interacting with Standard Model particles
has interactions with new vector bosons in precisely this mass range.  The proposed
experiment can probe these hypothetical particles with a sensitivity
that is un-rivaled by any existing or planned experiment.


\subsection*{Acknowledgements}

We thank Nima Arkani-Hamed, James Bjorken, Douglas Finkbeiner, Mathew Graham, John Jaros, 
Takashi Maruyama, 
Ken Moffeit, Richard Partridge, Michael Peskin, Maxim Pospelov, Allen Odian, Art Snyder, 
Jay Wacker, and Neal Weiner for many useful discussions 
and encouragement.  We also thank Johan Alwall for answering questions about Madgraph.  
We especially thank Peter Bosted, Kees de Jager, Doug Higinbotham, John LeRose, and Yi Qiang for their 
help in preparing the proposal for JLab Hall A.  RE and PS are supported by the US DOE under contract number 
DE-AC02-76SF00515.  RE, PS, and NT would like to thank the Kavli Institute for Theoretical Physics in 
Santa Barbara for hospitality during part of this research.  PS and NT also thank the Perimeter Institute 
for Theoretical Physics for hospitality.  
We also thank the SLAC National Accelerator Laboratory for providing funding for 
the ``Dark Forces'' workshop in September 2009.

\appendix

\section{Effective Photon Flux, Target Nucleus and Beam-Energy Dependence}\label{sec:chi}
In this appendix we summarize the formulas used in Section \ref{sec:reaction} for the reduced effective photon flux $\tilde \chi$, and highlight its dependence on the $A'$ mass, target nucleus, and beam energy.  
 The effective photon flux $\chi$ is obtained as in \cite{Kim:1973he,Tsai:1973py} by integrating electromagnetic form-factors over allowed photon virtualities:
 
For a general electric form factor $G_2(t)$,
\be
\chi \equiv \int_{t_{min}}^{t_{max}}  dt \f{t-t_{min}}{t^2} G_2(t) \label{ChiExp}
\ee
(the other form factor, $G_1(t)$, contributes only a negligible amount in 
all cases of interest).   Since we are dominated by a coherent scattering with $G_2 \propto Z^2$, it is useful to define a reduced photon flux,
\be
\tilde \chi \equiv \chi/Z^2. \label{ChiTilde}
\ee
The integral in \eqref{ChiExp} receives equal contributions at all $t$, and so is logarithmically sensitive to $t_{min}=(m_{A'}^2/2E_0)^2$ and $t_{max}=m_{A'}^2$.

For most energies in question, $G_2(t)$ is dominated by an elastic component 
\be
G_{2,el}(t)= \left(\f{a^2 t}{1+a^2 t} \right)^2
\left(\f{1}{1+t/d} \right)^2 Z^2,
\ee
where the first term parametrizes electron screening (the elastic atomic form factor) 
with $a=111\,Z^{-1/3}/m_e$, and the second finite nuclear size (the elastic nuclear form 
factor) with $d=0.164 \mbox{ GeV}^2 A^{-2/3}$.  
We have multiplied together the simple parametrizations used for each in \cite{Kim:1973he}.  
The logarithm from integrating \eqref{ChiExp} is large for 
$t_{min} < d$, which is true for most of the range of interest.  
However, for heavy $A'$, the elastic contribution is suppressed and is 
comparable to an inelastic term,
\be
G_{2, in}(t)= \left(\f{a'^2 t}{1+a'^2 t} \right)^2 \left(\f{1+\f{t}{4
    m_p^2} (\mu_p^2-1)}{(1+\f{t}{0.71\,{\rm GeV}^2})^4} \right)^2 Z,
\ee
where the first term parametrizes the inelastic atomic form factor and the second 
the inelastic nuclear form factor, and where $a'=773 \,Z^{-2/3}/m_e$, $m_p$ is the proton 
mass, and $\mu_p=2.79$ \cite{Kim:1973he}.  
This expression is valid when $t/4m_p^2$ is small, which is the case for $m_{A'}$ 
in the range of interest in this paper.  At large $t$ the form factors will deviate from these simple parameterizations but can be measured from data.  One can show that the contribution from the other inelastic nuclear form factor $G_1(t)$ is negligible.   

\begin{figure}
\includegraphics[width=0.455\textwidth]{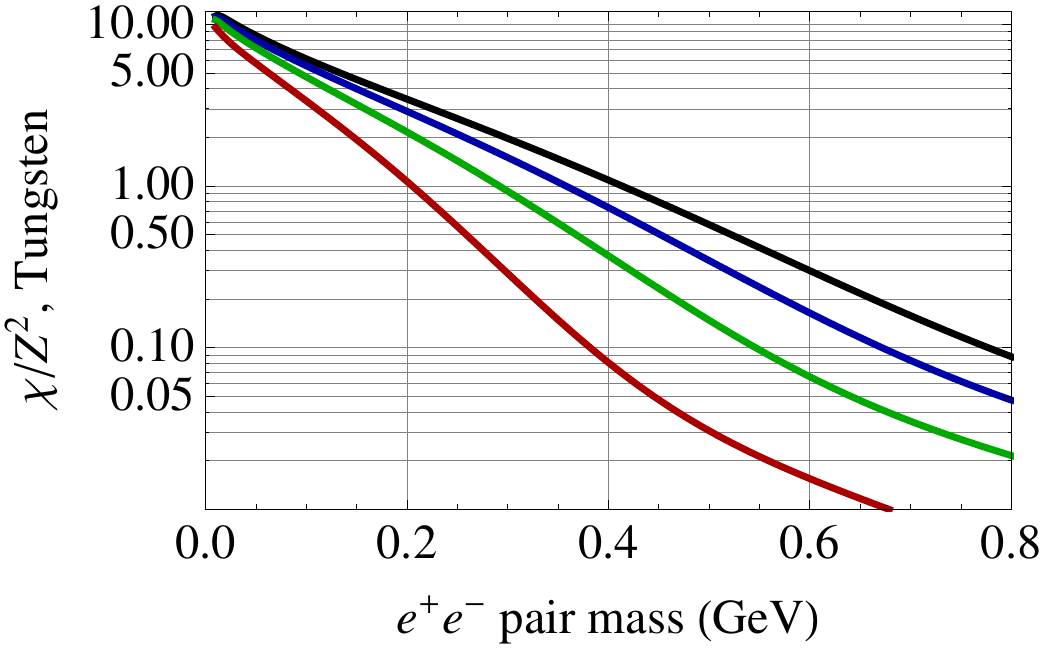}
\includegraphics[width=0.445\textwidth]{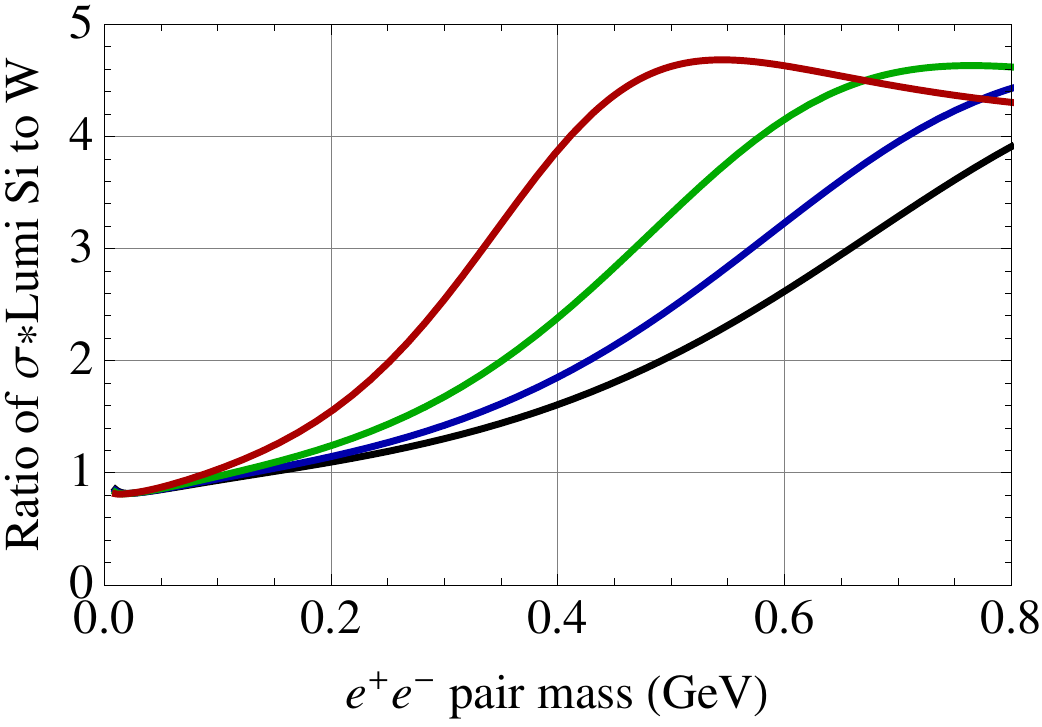}
\caption{\label{fig:chiLOG} \emph{Top:} The factor $\tilde \chi = \chi/Z^2$ defined in \eqref{ChiTilde} and \eqref{ChiExp} as a function of $\epm$ mass for (bottom to top) 1, 2, 3, and 4 GeV incident electrons on a Tungsten target.  \emph{Bottom:} The ratio of \eqref{ProductionRatio} $A'$ production rates per radiation length for Silicon and Tungsten targets, as a function of invariant mass and for beam energies (top to bottom at 0.4 GeV) 1, 2, 3, and 4 GeV incident electrons.}
\end{figure}

The resulting reduced form factor $\tilde \chi(m^2,E_0)=\chi/Z^2$ are plotted in the left panel of Figure \ref{fig:chiLOG} as a function of $e^+e^-$ mass for various electron energies (1, 2, 3, and 4 GeV) incident on a Tungsten target.  The relative efficiency of $A'$ production in targets of different compositions but the same thickness in radiation lengths is given by the ratio 
\be
R(Z_1,Z_2)=\frac{X_0(Z_1) \chi(Z_1,t)/A(Z_1)}{X_0(Z_2) \chi(Z_2,t)/A(Z_2)}.\label{ProductionRatio}
\ee
For example the ratio $R(Si,W)$ is shown in the right panel  of Figure \ref{fig:chiLOG}, again as a function of $e^+e^-$ mass for beam energies between 1 and 4 GeV.

\section{Mass resolution}
\label{sec:resolution}

In this appendix, we briefly describe an estimate of the mass resolution of the spectrometer.  
Since we are looking for a small bump on the invariant mass spectrum distribution, an excellent 
mass resolution is essential to obtain a good reach in $\epsilon$.   

The mass resolution of the spectrometer, $\delta_m$, is roughly given by
\be
\left(\frac{\delta_m}{m}\right)^2 = \left(\frac{\delta_{p}}{p}\right)^2 + 
0.5 \times \left(\frac{\delta_{\theta}}{\theta}\right)^2,
\ee
where $\delta_\theta$ is the angular resolution of the electron or positron, 
and $\delta_p/p$ is the  momentum resolution of the HRS, which is always less 
than $3 \times 10^{-4}$ (in our estimates for the reach of $\epsilon$, we take $\delta p/p$ to be 
equal to this upper bound).  
We have
\be
(\delta_{\theta})^2 = (\delta_{{HRS}})^2 + (\delta_{\theta}^{ms})^2,
\ee
where $\delta_{HRS}$ is the HRS angular resolution, which is $\sim 0.5$ mrad in the 
horizontal direction and $\sim 1$ mrad in the vertical direction.  
Moreover, $\delta_\theta^{ms}$ represents the degradation of the resolution due to 
multiple Coulomb scattering in the target.  It is given by the standard formula \cite{Amsler:2008zzb}
\be
\delta_{\theta}^{ms} = \f{13.6}{p[{\rm MeV}]} \,\sqrt{\f{t}{X_0}} \,\left[1+0.038 \ln \Big(\f{t}{X_0}\Big)\right],
\ee
where $t$ is the thickness in radiation lengths of the material along the path of the particle,
$X_0$ is the radiation length of the target in g/cm$^2$, and $p$ is the momentum of 
particle in MeV.

For the proposed experiment, the thickness of the target along the
direction of the beam line varies from $t = 0.003 X_0$ to $t=0.09
X_0$.  However, in the case of a foil target or a target composed of
several thin wires, the distance traversed by trident
electron-positron pairs can be significantly smaller because the
electron and positron have relatively large angles with respect to the
beam line.  For a foil, we can take $t \approx \frac{1}{2} t_f
/\sin(\theta_p-\theta_f)$, where $t_f$ is the foil thickness and
$\theta_f$ its angle relative to the beam line.  In this case the
effective thickness traversed by the beam is $T_0 = t_f/\sin\theta_f$.

In the case of a target composed of multiple wires, as was assumed in
determining the experimental sensitivity, the wires can be spaced
widely enough that the pair-produced particles need only travel
through a single wire.  In this case, $t$ is typically the radius of
the wire, which for Tungsten wire targets can be as small as $10 \mu$m, or $3 \times 10^{-3} X_0$.  In this case, we find that the HRS angular
resolution, $\delta_{HRS}$, is comparable to the multiple scattering
$\delta_\theta^{ms}$ in the proposed experiment.

\section{Monte Carlo validation with E04-012 data}
\label{sec:validation}

In this appendix, we briefly describe a validation of the Monte Carlo (MC) simulation of the 
signal, Bethe-Heitler and radiative trident backgrounds (shown in Figure \ref{fig:diagrams} and discussed in 
\S \ref{sec:signalTrident}), and the positron singles.  

We first discuss a comparison of the MC with previous experimental results from the JLab experiment E04-012 
\cite{E04-012}.  
This experiment consisted of a 5.01 GeV, $14.5 \mu A$ electron beam 
incident on a 
1.72\% radiation length liquid Hydrogen target.  
The $e^+$ singles rate was measured to be $\sim 1.1$ kHz 
in a momentum window of 
 $\pm 4\%$ around 1.93 GeV and an angular acceptance of 4.5 msr with an aspect ratio of 2-to-1
centered at an angle of $6^\circ$.  
The $e^+e^-$ coincidence rate was measured at $\sim 4$ Hz for the same angular acceptance 
for both the electron and positron arm, and with a momentum window of $\pm 4\%$ around 1.93 GeV  
for the positron and $\pm 4\%$ around 1.98 GeV for the electron.
We simulated this with MadGraph and MadEvent~\cite{Alwall:2007st} as 
described in \S \ref{sec:signalTrident}, using a form factor for Hydrogen given in 
\cite{Kim:1973he}.
We find a $e^+$ singles rate of $\sim 965$ Hz  and an $e^+e^-$ coincidence rate of 3.9 Hz, which agrees 
with the measured rates to within $\sim 19\%$ and a few percent, respectively.  

We have also verified the implementation of form factors in Monte Carlo by simulating photo-production of electrons and muons off Tungsten and Beryllium with MadGraph and MadEvent.  The resulting cross-sections agree to within $30\%$ with published computations in the  Weizs\"acker-Williams approximation~\cite{Tsai:1973py}.

\bibliography{ArxivProposal}
\bibliographystyle{JHEP}

\end{document}